\documentclass[a4paper,11pt]{article}

\usepackage{jheppub}
\usepackage[table]{xcolor}
\usepackage{slashbox}
\allowdisplaybreaks
\usepackage{graphicx}
\usepackage{epstopdf}
\usepackage{tikz}

\newcommand{\be}{\begin{equation}}
\newcommand{\ee}{\end{equation}}
\newcommand{\bea}{\begin{eqnarray}}
\newcommand{\eea}{\end{eqnarray}}
\newcommand{\nn}{\nonumber}

\newcommand{\bdm}{\begin{displaymath}}
\newcommand{\edm}{\end{displaymath}}

\title{NLO gravitational quartic-in-spin interaction}

\author[a]{Mich\`ele Levi,}
\author[b]{Fei Teng}

\affiliation[a]{Niels Bohr International Academy, Niels Bohr Institute,
University of Copenhagen,
\\Blegdamsvej 17, 2100 Copenhagen, Denmark}

\affiliation[b]{Department of Physics and Astronomy, Uppsala University, 
75108 Uppsala, Sweden}

\emailAdd{michelelevi@nbi.ku.dk, fei.teng@physics.uu.se}

\preprint{UUITP--32/20}

\abstract{In this paper we derive for the first time the complete gravitational 
quartic-in-spin interaction of generic compact binaries at the next-to-leading order 
in the post-Newtonian (PN) expansion. 
The derivation builds on the effective field theory for gravitating spinning objects, 
and its recent extensions, in which new type of worldline couplings should be considered, 
as well as on the extension of the effective action of a spinning particle to quadratic 
order in the curvature. The latter extension entails a new Wilson coefficient that 
appears in this sector.
This work pushes the precision frontier with spins at the fifth PN (5PN) order 
for maximally-spinning compact objects, and at the same time informs us of the 
gravitational Compton scattering with higher spins.}

\begin{document}

\maketitle

\flushbottom

\section{Introduction} 
\label{intro}

In this work we pursue two interwoven objectives, each of which has a different nature. 
The first objective falls under the timely phenomenological efforts to improve the 
theoretical precision for modeling gravitational waveforms, that are currently 
successfully being used to measure gravitational waves (GWs) from mergers of compact binaries 
as of 2016 
\cite{Abbott:2016blz}. The post-Newtonian (PN) approximation of classical Gravity 
\cite{Blanchet:2013haa} via the effective-one body approach \cite{Buonanno:1998gg} provides 
the crucial input for the generation of GW templates, and accordingly significant progress 
has been carried out recently in pushing the state of the art, with the current precision 
frontier at the fourth and a half PN (4.5PN) 
\cite{Levi:2019kgk,Levi:2020kvb,Antonelli:2020aeb} and the 5PN 
\cite{Foffa:2019hrb,Blumlein:2019zku,Foffa:2019eeb,Levi:2020uwu,Bini:2020wpo,Bini:2020uiq}
orders in the accuracy of the orbital dynamics.
The complete state of the art of the generic orbital dynamics of a compact binary to date is 
captured in table \ref{stateoftheart}.
In order to complete a certain PN accuracy, the sectors that are shown in table 
\ref{stateoftheart} should be addressed across its diagonals, which stands for two orthogonal 
types of challenges: The challenge of going along the horizontal axis of the table is that of 
computational multi-loop technology, whereas the challenge of going along the vertical axis 
of the table involves improving the conceptual understanding of spin in gravity and 
higher-spin theory. 

\begin{table}[t]
\begin{center}
\begin{tabular}{|l|r|r|r|r|r}
\hline
\backslashbox{\quad\boldmath{$l$}}{\boldmath{$n$}} 
&  (N\boldmath{$^{0}$})LO
& N\boldmath{$^{(1)}$}LO & \boldmath{N$^2$LO}
& \boldmath{N$^3$LO} & \boldmath{N$^4$LO}\\
\hline
\boldmath{S$^0$} & 1 & 0  & 3 
& 0  & 25 \\
\hline
\boldmath{S$^1$} & 2 & 7  & 32 
& \textbf{174} 
& \\
\hline
\boldmath{S$^2$} & 2 & 2  & \textbf{18} 
& \textbf{52} 
& \\ 
\hline
\boldmath{S$^3$} & 4  & \cellcolor[gray]{0.9} \textbf{24}
& \cellcolor[gray]{0.9} & \cellcolor[gray]{0.9} & \cellcolor[gray]{0.9}\\
\hline
\boldmath{S$^4$} & 3 & \cellcolor[gray]{0.9} \textbf{5} & \cellcolor[gray]{0.9} 
& \cellcolor[gray]{0.9} & \cellcolor[gray]{0.9}\\
\end{tabular}
\caption{The complete state of the art of PN orbital dynamics of generic compact binaries. 
The PN corrections enter at the order $n + l + \text{Parity}(l)/2$, with the parity $0$ or 
$1$ for even or odd $l$, respectively. The higher-order sectors with the entries in boldface 
have been uniquely addressed in \cite{Levi:2015ixa,Levi:2019kgk,Levi:2020kvb,Levi:2020uwu} 
and in the present 
work via the EFT of gravitating spinning objects \cite{Levi:2015msa}. The entries in the 
table indicate the loop computational scale within this framework, namely the number of 
highest-loop graphs in each sector. The sectors up to the current complete state of the 
art at the 4PN order, except the top right one, are available in the public {\tt{EFTofPNG}} 
code at \url{https://github.com/miche-levi/pncbc-eftofpng} \cite{Levi:2017kzq}.} 
\label{stateoftheart}
\end{center}
\end{table}

\begin{figure}[t]
\centering
\includegraphics[scale=1]{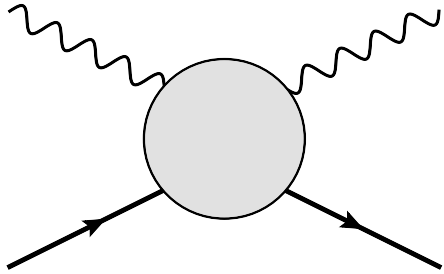}
\caption{The gravitational Compton scattering with two massive spin 
particles and two gravitons.}
\label{compton}
\end{figure}

This brings us to the second objective of the present work, which is the theoretical effort 
to understand what happens in gravitational interactions when higher spins are involved.
Within classical gravity the first work to provide the LO quadratic-in-spin interaction, 
including the spin-induced quadrupole was carried out decades ago \cite{Barker:1975ae}. 
This was considered much later
within the effective field theory (EFT) approach to PN Gravity \cite{Goldberger:2004jt}, 
whose extension to the spinning case was first approached in \cite{Porto:2005ac}, with 
effective parameters referred to as Wilson coefficients, similar to the spin-induced 
quadrupolar 
deformation parameter in \cite{Poisson:1997ha}. The EFT of gravitating spinning objects was  
then introduced in \cite{Levi:2015msa}, where the tower of gravitational couplings to 
all orders in spin, that are linear in the curvature, was provided. Later, a new 
spinor-helicity formalism for massive particles of any spin was introduced in 
\cite{Arkani-Hamed:2017jhn}, and was instrumental to recent studies within the language of 
scattering amplitudes of classical spin effects
\cite{Guevara:2017csg,Guevara:2018wpp,Chung:2018kqs,Chung:2019duq,Aoude:2020onz,Chung:2020rrz,
Bern:2020buy}. Scattering amplitudes involving a quantum spin of $s=l/2$ correspond to 
classical interactions
with spin to the $l$th order, and hence the gray area in table \ref{stateoftheart} 
corresponds to where the gravitational Compton scattering (see figure \ref{compton}) with a 
quantum spin $s > 1$ is required, as of the one-loop level. The fundamental issue is that the 
gravitational Compton amplitude cannot be uniquely fixed for $s>2$, related with the tension 
in formulating a perturbative UV completion of gravity with higher spins~\cite{Arkani-Hamed:2017jhn}. 

Interestingly, at the classical level the gray area in table \ref{stateoftheart} also 
corresponds to where the linear momentum can no longer be taken as independent of the spin, \
and new types of complicate contributions arise in the interactions, as already pointed out 
in \cite{Levi:2015msa}. At this stage it is not clear whether one gets a unique well-defined 
classical result, which resonates with the difficulty of uniquely fixing the graviton Compton 
amplitude with higher spins, and thus these two inquiries may inform each other.

This work directly builds on the EFT of gravitating spinning  objects introduced in 
\cite{Levi:2015msa}, its implementation on \cite{Levi:2014gsa} at LO, and the recent 
extension to the cubic-in-spin interaction at NLO in \cite{Levi:2019kgk}, as well as on the 
recent study of the NNNLO quadratic-in-spin interaction in \cite{Levi:2020uwu}, to get 
the quartic-in-spin interaction at the NLO. This sector enters at the 5PN order for 
maximally-spinning compact objects, beyond the current complete PN state of the art in 
general, and with spins in particular \cite{Levi:2016ofk}. This paper, that pushes PN 
precision with higher spin, extends the body of work of various past studies in  
\cite{Hergt:2007ha,Porto:2008jj,Steinhoff:2008ji,Hergt:2008jn,Hergt:2010pa,Levi:2014gsa,
Vaidya:2014kza,Marsat:2014xea,Levi:2015msa,Levi:2015ixa,Vines:2016qwa,
Levi:2019kgk,Levi:2020uwu}, where here we derive the complete quartic-in-spin sector, 
including all interactions 
with all possible spin-induced higher-order multipoles up to and including the hexadecapole. 
This sector is also the next complicate one after that in \cite{Levi:2019kgk} in the 
intriguing gray area of table 
\ref{stateoftheart}, and thus beyond pushing the state of the art in PN theory, we should be 
alert to the new conceptual features that show up in it.

The paper is organized as follows. In section \ref{theory} we review briefly
the formulation of the EFT of spinning gravitating objects from \cite{Levi:2015msa}, 
and outline the ingredients relevant to this sector. Notably, in section \ref{moretheory} we 
extend the effective theory of a spinning particle, in order to extract new contributions to 
the current sector. In section 
\ref{Feyncompute} we present the full perturbative expansion with the elementary worldline 
couplings of the EFT, where the linear momentum is taken as independent of the spin, and in 
section \ref{newfromgauge?} we handle new contributions due to composite worldline 
couplings that emerge, similar to what was first seen in the NLO cubic-in-spin sector 
\cite{Levi:2019kgk}. In 
section \ref{quadcurv} we consider further new contributions arising from new worldline 
couplings that are quadratic in the curvature. The final result for the total effective 
action of the complete NLO gravitational quartic-in-spin
sector is provided in section \ref{finalresult}, and we conclude the work in section 
\ref{lafin}. An ancillary file of Mathematica notebook that contains all the results of this 
paper is included in the arXiv submission.

\section{The EFT of gravitating spinning objects}
\label{theory}

The evaluation of this sector, that contains spins at quartic order with leading 
gravitational nonlinearities, builds on the EFT of gravitating spinning objects formulated in 
\cite{Levi:2015msa}, its implementation from LO to NNLO up to the complete state of the art 
at the 4PN order accomplished in \cite{Levi:2015msa,Levi:2015uxa,Levi:2015ixa}, the LO 
complete cubic- and quartic-in-spin sectors obtained in \cite{Levi:2014gsa}, as well as on 
the recent complete NLO cubic-in-spin sector obtained in \cite{Levi:2019kgk} at the 4.5PN 
order, and the recent studies of the NNNLO spin-orbit and quadratic-in-spin sectors in 
\cite{Levi:2020kvb,Levi:2020uwu} at 4.5 and 5PN orders. 
We follow all the conventions and gauge 
choices in the aforementioned papers, and we also use the beneficial Kaluza-Klein (KK) 
decomposition of the field to scalar, vector and tensor components, as in all high-order PN 
computations in the EFT approach \cite{Kol:2007bc,Kol:2010ze,Levi:2018nxp}.

We start from a two-particle effective action \cite{Levi:2018nxp}, in which each of the 
objects is captured by the one-particle effective action of a spinning particle, that was 
derived in \cite{Levi:2015msa}. This effective action contains a pure gravitational piece, 
and two copies of a point-particle action that captures the coupling of gravity to the 
worldline degrees of freedom (DOFs). The Feynman rules for the propagators and 
self-interaction vertices derived from the purely gravitational action are found in 
\cite{Levi:2011eq} and \cite{Levi:2015uxa}. The effective action of a spinning particle, 
including its spin-induced non-minimal coupling, and in which the gauge freedom of the 
rotational DOFs was introduced into the action, has the form \cite{Levi:2015msa}
\begin{align} \label{spinact}
S_{\text{pp}}(\sigma)=&\int d\sigma\left[-m \sqrt{u^2}
-\frac{1}{2} \hat{S}_{\mu\nu} \hat{\Omega}^{\mu\nu}
-\frac{\hat{S}^{\mu\nu} p_{\nu}}{p^2} 
\frac{D p_{\mu}}{D \sigma} + L_{\text{SI}}\right],
\end{align}
with the four-velocity $u^{\mu}$, the conjugate linear momentum $p_\mu$ and the generic 
rotational DOFs, denoted with a hat, e.g.~$\hat{S}_{\mu\nu}$. The spin-induced part, labeled 
by ``SI'', contributes to this sector via its three leading terms \cite{Levi:2015msa}:
\begin{align} \label{nmc}
L_{\text{SI}} =&
-\frac{C_{ES^2}}{2m} \frac{E_{\mu\nu}}{\sqrt{u^2}} S^\mu S^\nu
-\frac{C_{BS^3}}{6m^2}\frac{D_{\lambda}B_{\mu\nu}}{\sqrt{u^2}}
S^{\mu} S^{\nu}S^\lambda
+\frac{C_{ES^4}}{24m^3} \frac{D_{\kappa}D_{\lambda}E_{\mu\nu}}{\sqrt{u^2}} 
S^\mu S^\nu S^\lambda S^\kappa,
\end{align}
in which we use the electric and magnetic components of the curvature tensor, with the 
Levi-Civita tensor density in curved spacetime in the latter, for definite parity
\bea
E_{\mu\nu}&\equiv& R_{\mu\alpha\nu\beta}u^{\alpha}u^{\beta}, \label{eq:E}\\
B_{\mu\nu}&\equiv& \frac{1}{2} \epsilon_{\alpha\beta\gamma\mu} 
R^{\alpha\beta}_{\,\,\,\,\,\,\,\delta\nu}u^{\gamma}u^{\delta}\label{eq:B},
\eea
of even and odd parity, respectively, and a classical version of the Pauli-Lubanski 
pseudovector, $S^{\mu}$, as defined first in \cite{Levi:2014gsa}:
\begin{align}
S_{\mu}=\frac{1}{2}\epsilon_{\alpha\beta\gamma\mu}S^{\alpha\beta}\frac{p^{\gamma}}{p}.
\end{align}
Note that this definition is with a reverse sign with respect to the one detailed in  
\cite{Levi:2015msa}, which was implemented there up to the quadratic-in-spin order, where the 
sign choice does not make a difference. Further, note that $S^{\alpha\beta}$ here is not the 
generic spin variable, but rather its projection to the rest frame, see eq.~(3.29) in 
\cite{Levi:2015msa}

We remind the extra term with a covariant derivative of the linear momentum in 
eq.~\eqref{spinact}, which is essentially the Thomas precession, that was derived from the 
introduction of gauge freedom to the rotational DOFs in \cite{Levi:2015msa} (later recovered 
in e.g.~\cite{Chung:2019duq} as ``Hilbert space matching''). This term is relevant to all 
orders in spin, including all finite size spin 
effects, though it does not encapsulate any UV physics, but rather accounts for the extended 
measure of a relativistic gravitating spinning object.

Since we compute here the \textit{complete} NLO quartic-in-spin sector our graphs  
contain all spin-induced higher-multipoles up to the hexadecapole, in addition to the mass 
and spin. Therefore we need the Feynman rules of worldline-graviton coupling to NLO for all 
of these multipoles, including new rules for the hexadecapole couplings. The Feynman rules 
for the mass, the spin and the spin-induced quadrupole are found e.g.~in \cite{Levi:2010zu}, 
\cite{Levi:2015msa}, \cite{Levi:2015uxa}, and \cite{Levi:2015ixa}. The spin-induced 
octupole couplings are found in \cite{Levi:2014gsa,Levi:2019kgk}.
In this work the Feynman rule of the scalar component of the KK fields, which appeared at LO 
in \cite{Levi:2014gsa}, should be extended to the next PN order and is given as follows:
\begin{align}
\label{eq:s4phi}   \parbox{12mm}{\includegraphics[scale=0.6]{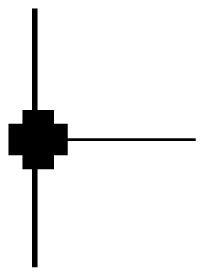}}
 =& \int dt \,\,-\frac{C_{\text{ES}^4}}{24m^3}\Bigg[S_{i}S_{j}S_{k}S_{l}
  \left(\partial_i\partial_j\partial_k\partial_l\phi\left(1+\frac{3}{2}v^2\right)
  -2v^iv^m\partial_m\partial_j\partial_k\partial_l\phi\right)\nonumber\\
  &\qquad\qquad\qquad +S^2S_jS_k
  \left(v^mv^l\partial_m\partial_l\partial_j\partial_k\phi
  +2v^l\partial_l\partial_j\partial_k\partial_t\phi
  +\partial_j\partial_k\partial_t^2\phi\right) \Bigg],
\end{align}
where the crossed black box represents the spin-induced hexadecapole.
In this Feynman rule and all the rules that follow in this paper, all indices are Euclidean, 
the spin is fixed in the canonical gauge in the local frame, and 
$S_i\equiv\epsilon_{ijk}S_{jk}$ (for more details on these points, see \cite{Levi:2015msa}). 
We also need new rules for the one-graviton coupling of the KK vector field, and for the 
two-graviton coupling with the KK scalar field, which are given as follows:
\begin{align}
\label{eq:s4A} 
\parbox{12mm}{\includegraphics[scale=0.6]{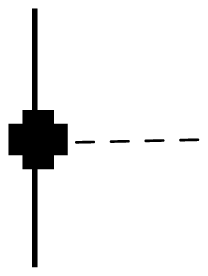}}
 =&\int dt \,\,\frac{C_{\text{ES}^4}}{24m^3}S_iS_jS_kS_l
 \left[ v^m\partial_i\partial_j\partial_k\partial_l A_m
 -v^m\partial_m\partial_j\partial_k\partial_l A_i
 -\partial_j\partial_k\partial_l\partial_t A_i\right],
 \end{align}
\begin{align}
\label{eq:s4phi2}  \parbox{12mm}{\includegraphics[scale=0.6]{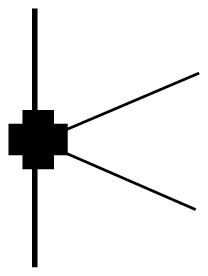}}
 =& \int dt \,\, -\frac{C_{\text{ES}^4}}{24m^3}\Bigg[ 
 S_iS_jS_kS_l\Big(16\partial_i\partial_j\partial_k\phi\partial_l\phi
 +10\partial_i\partial_j\phi\partial_k\partial_l\phi
 +5\phi\partial_i\partial_j\partial_k\partial_l\phi\Big)\nonumber\\
 &\qquad\qquad\qquad-S^2S_jS_k\Big(7\partial_j\partial_k\partial_n\phi\partial_n\phi
 +4\partial_j\partial_n\phi\partial_k\partial_n\phi\Big)\Bigg].
\end{align}
Note the dominant role that the KK scalar field plays in the coupling to the even-parity 
hexadecapole, similar to the couplings to the even-parity mass monopole, and spin-induced 
quadrupole.

For this sector we also need to extend the non-minimal coupling part of the spinning particle 
action, and add higher-dimensional operators beyond the linear-in-Riemann ones, which were 
derived to all orders in spin in \cite{Levi:2015msa}, but we also need to address the new 
feature that became relevant as of the NLO cubic-in-spin sector (the simplest corner of the 
gray area in table~\ref{stateoftheart}) \cite{Levi:2019kgk}. For the latter we need to take 
into account corrections to the linear momentum that are linear in Riemann and higher-order 
in the spin similar to \cite{Levi:2019kgk}, which was already explicitly noted in 
\cite{Levi:2015msa}. These corrections give rise to what we will refer to as ``composite'' 
worldline couplings, which are considered in section \ref{newfromgauge?}, after the 
``elementary'' worldline couplings, in which the linear momentum is still independent of the 
spin (the white area of table \ref{stateoftheart}), are covered in section~\ref{Feyncompute}.

\subsection{Extending the EFT of a spinning particle}
\label{moretheory}

As noted above at the 5PN order the effective action, or 
more specifically, the non-minimal coupling part of the effective action of a spinning 
particle, that is given in eq.~\eqref{nmc}, should be extended as done in 
\cite{Levi:2020uwu}. This extension should include operators that are 
higher-order in the curvature components, and describe the tidal deformations of the compact 
object. This extension should only include operators that are quadratic in the curvature,  
since here only corrections up to order $G^2$ are considered. Following the symmetries and 
the logic detailed in \cite{Levi:2015msa}, and implemented in the recent extension in 
\cite{Levi:2020uwu}, 
we find that the new terms to quartic order in spin are of the form:
\begin{align} \label{nmcspinRR}
L_{\text{NMC(R$^2$)}}&
= C_{E^2} \frac{E_{\alpha\beta}E^{\alpha\beta}}{\sqrt{u^2}^{\,3}}
+ C_{B^2} \frac{B_{\alpha\beta}B^{\alpha\beta}}{\sqrt{u^2}^{\,3}}+\ldots
\nn\\&
+ C_{E^2S^2} S^{\mu} S^{\nu} \frac{E_{\mu\alpha}E_{\nu}^{\,\alpha}}{\sqrt{u^2}^{\,3}}
+ C_{B^2S^2} S^{\mu} S^{\nu} \frac{B_{\mu\alpha}B_{\nu}^{\alpha}}{\sqrt{u^2}^{\,3}}
\nn\\&
+ C_{E^2S^4} S^{\mu} S^{\nu} S^{\kappa} S^{\rho} 
\frac{E_{\mu\nu}E_{\kappa\rho}}{\sqrt{u^2}^{\,3}}
+ C_{B^2S^4} S^{\mu} S^{\nu} S^{\kappa} S^{\rho}
\frac{B_{\mu\nu}B_{\kappa\rho}}{\sqrt{u^2}^{\,3}}
\nn\\&
+ C_{\nabla EBS} S^{\mu} \frac{D_{\mu}E_{\alpha\beta}B^{\alpha\beta}}{\sqrt{u^2}^{\,3}}
+ C_{E\nabla BS} S^{\mu} \frac{E_{\alpha\beta}D_{\mu}B^{\alpha\beta}}{\sqrt{u^2}^{\,3}}
\nn\\& 
+ C_{\nabla EBS^3} S^{\mu} S^{\nu} S^{\kappa} 
\frac{D_{\kappa}E_{\mu\alpha}B_{\nu}^{\alpha}}{\sqrt{u^2}^{\,3}}
+ C_{E\nabla BS^3} S^{\mu} S^{\nu} S^{\kappa}
\frac{E_{\mu\alpha}D_{\kappa}B_{\nu}^{\alpha}}{\sqrt{u^2}^{\,3}}
\nn\\& 
+ C_{(\nabla E)^2S^2} S^{\mu} S^{\nu} 
\frac{D_{\mu}E_{\alpha\beta}D_{\nu}E^{\alpha\beta}}{\sqrt{u^2}^{\,3}}
+ C_{(\nabla B)^2S^2} S^{\mu} S^{\nu} 
\frac{D_{\mu}B_{\alpha\beta}D_{\nu}B^{\alpha\beta}}{\sqrt{u^2}^{\,3}}
\nn\\& 
+ C_{(\nabla E)^2S^4} S^{\mu} S^{\nu} S^{\kappa} S^{\rho}
\frac{D_{\kappa}E_{\mu\alpha}D_{\rho}E_{\nu}^{\alpha}}{\sqrt{u^2}^{\,3}}
+ C_{(\nabla B)^2S^4} S^{\mu} S^{\nu} S^{\kappa} S^{\rho}
\frac{D_{\kappa}B_{\mu\alpha}D_{\rho}B_{\nu}^{\alpha}}{\sqrt{u^2}^{\,3}}+\ldots, 
\end{align}
with new tidal Wilson coefficients involved with these new operators, which in contrast to 
those in eq.~\eqref{nmc} -- are currently defined to absorb all numerical and 
mass factors. 

As in \cite{Levi:2020uwu} in the first line of eq.~\eqref{nmcspinRR} the leading mass-induced 
quadrupolar tidal deformations, which are known to enter at the 5PN order are written 
\cite{Levi:2018nxp}, suppressing higher-order mass-induced tidal operators, that were 
provided in \cite{Bini:2012gu}. We have then written the adiabatic tidal operators with spin 
up to the quartic order. From power-counting considerations, which were detailed in 
\cite{Levi:2015msa,Levi:2018nxp}, it can be deduced that the second and third lines of 
eq.~\eqref{nmcspinRR} contribute as of the 5PN order, whereas the fourth and fifth lines, and 
the sixth and seventh lines, of eq.~\eqref{nmcspinRR}, contribute as of the 6.5PN and 7PN 
orders, respectively. 
Therefore at the 5PN order considered in this work 
there are two new operators which are quartic in the spin and quadratic in the curvature 
that are given by
\begin{align} \label{tidalS4}
L_{\text{S$^4$(R$^2$)}}&
=  \frac{C_{E^2S^4}}{24m^3} S^{\mu} S^{\nu} S^{\kappa} S^{\rho} 
\frac{E_{\mu\nu}E_{\kappa\rho}}{\sqrt{u^2}^{\,3}}
+ \frac{C_{B^2S^4}}{24m^3} S^{\mu} S^{\nu} S^{\kappa} S^{\rho}
\frac{B_{\mu\nu}B_{\kappa\rho}}{\sqrt{u^2}^{\,3}},
\end{align}
in which the numerical and mass factors have now been adjusted to render the Wilson 
coefficients dimensionless. 

The consequent new Feynman rule that contributes from eq.~\eqref{tidalS4} is then a 
two-graviton coupling of the KK scalar field given by 
\begin{align}
\label{frs4E2phi2} \parbox{12mm}{\includegraphics[scale=0.6]{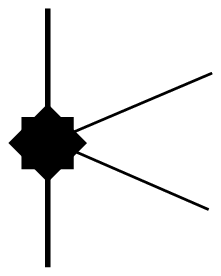}}
 =& \int dt \,\, \frac{C_{\text{E}^2\text{S}^4}}{24m^3}\Bigg[ 
 S_iS_jS_kS_l\phi_{,ij}\phi_{,kl}
 -2S^2S_iS_j\phi_{,ij}\phi_{,kk}+S^4\phi_{,ii}\phi_{,jj}\Bigg],
\end{align}
that arises from the quadratic electric operator, and is represented here by the black ``star 
of Lakshmi''. As to the quadratic magnetic operator in eq.~\eqref{tidalS4}, it contributes a 
two-graviton coupling of the KK vector field, which will become relevant only at the 6PN 
order, namely beyond the present sector.

\section{Elementary worldline couplings}
\label{Feyncompute}

Let us first evaluate the Feynman graphs from the diagrammatic expansion of the effective 
action, that contain the elementary worldline couplings, i.e.~those obtained under the 
leading approximation of the linear momentum that is independent of spin. With spins all of 
the three relevant topologies up to the $G^2$ order are realized even with the useful KK
decomposition \cite{Levi:2008nh,Levi:2010zu,Levi:2015msa,Levi:2018nxp}, and we have here a 
total of $23=10+8+5$ graphs, distributed among the topologies of one- and two-graviton 
exchanges, and cubic self-interaction, as shown in
figures \ref{s4nlo1g}-\ref{s4nlo1loop}  (drawn with 
Jaxodraw \cite{Binosi:2003yf,Binosi:2008ig} based on \cite{Vermaseren:1994je}), respectively.
As noted in table~\ref{stateoftheart} only 5 of the graphs require a one-loop evaluation, 
though as we consider the nonlinear graphs in the sector, the options to assemble a 
quartic-in-spin interaction multiply.

At the linear level there are only three types of interaction, similar to the LO in 
\cite{Levi:2014gsa}, namely either a quadrupole-quadrupole, an octupole-dipole
or a hexadecapole-monopole interaction. These graphs are easily constructed following the 
nice analogies among these types of interaction according to the parity of the multipole 
moments involved \cite{Levi:2014gsa,Levi:2019kgk}. Yet, more types of interaction emerge 
once we proceed to the nonlinear level, in which there are interactions with various 
multipoles on two different points of the worldline, which occurs as of the NLO spin-squared 
sector \cite{Levi:2015msa,Levi:2015ixa,Levi:2019kgk}, and which add up to an interaction that 
is quartic in the spin, such as a spin and a spin-induced quadrupole, or two spin-induced 
quadrupoles on the same worldline.

Note that all the graphs in this sector are to be included together
with their mirror images, in which the worldline labels are exchanged, $1\leftrightarrow2$, 
and that more details on the generation and the evaluation of the Feynman graphs can be found 
in \cite{Levi:2018nxp} and references therein.

\subsection{One-graviton exchange}

\begin{figure}[t]
\centering
\includegraphics[scale=0.5]{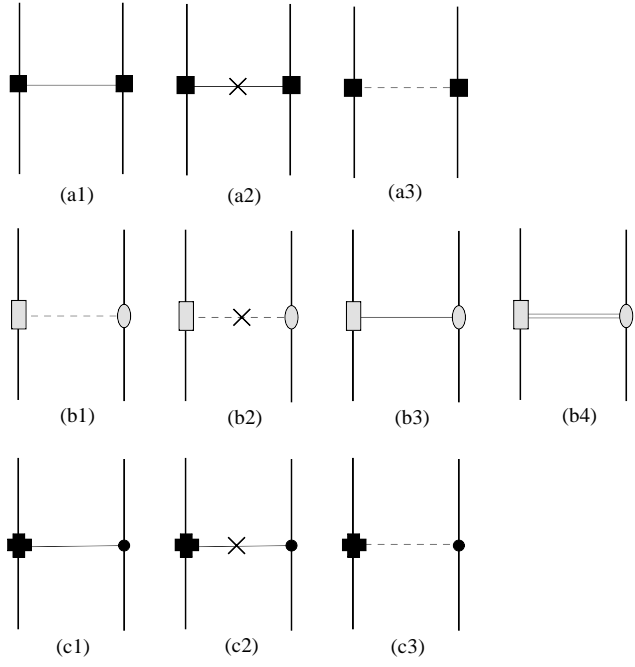}
\caption{The Feynman graphs of one-graviton exchange, that contribute to the NLO 
quartic-in-spin interaction at the 5PN order for maximally-rotating compact objects. 
At the linear level we only have three types of interaction, similar to the LO in 
\cite{Levi:2014gsa}, of either a quadrupole-quadrupole, octupole-dipole or a 
hexadecapole-monopole type. The graphs are easily constructed following the nice analogies 
pointed out in \cite{Levi:2014gsa} among the interactions according to the parity of the 
multipole moments involved. Notice that we have here the three graphs that appeared at the LO 
with the quadratic time insertions on the propagators at graphs (a2), (b2), (c2), and a
new hexadecapole coupling with the KK vector field at graph (c3).}
\label{s4nlo1g}
\end{figure}

There are $10$ graphs of one-graviton exchange in this sector, as seen in figure 
\ref{s4nlo1g}, in which most already contain time derivatives to be evaluated. 
As in previous works of one of the authors, all of the higher-order time derivative terms
that appear in the evaluations of the graphs are kept, to be treated rigorously later
via the procedure of redefinitions of variables, that was explained in detail in 
\cite{Levi:2014sba}.
Notice that there are 3 graphs that appeared at the LO, to which we add the
quadratic time insertions on the propagators at graphs 2(a2), (b2), (c2), and a
new hexadecapole coupling of the KK vector field at graph 2(c3).

The graphs in figure \ref{s4nlo1g} have the following values:
\begin{align}
\text{Fig.~2(a1)}&=\frac{3C_{1(\text{ES}^2)}C_{2(\text{ES}^2)}}{8}\frac{G}{m_1m_2r^5}\Big[
70 \big(\vec{S}_1\cdot \vec{n}\big)^2 \big(\vec{S}_2\cdot \vec{n}\big)^2\Big(1+\frac{3v_1^2}{2}+\frac{3v_2^2}{2}\Big)\nonumber\\
& -10 \big(\vec{S}_1\cdot \vec{n}\big)^2S_2^2\Big(1+\frac{3v_1^2}{2}+\frac{5v_2^2}{2}-7\big(\vec{v}_2\cdot \vec{n}\big)^2\Big) \nonumber\\
& -40 \big(\vec{S}_1\cdot\vec{n}\big) \big(\vec{S}_1\cdot \vec{S}_2\big) \big(\vec{S}_2\cdot \vec{n}\big)\Big(1+\frac{3v_1^2}{2}+\frac{3v_2^2}{2}\Big)\nonumber\\
& -70 \big(\vec{S}_1\cdot \vec{n}\big)^2\big(\vec{S}_2\cdot \vec{n}\big)\big(\vec{S}_2\cdot \vec{v}_2\big) \big(\vec{v}_2\cdot \vec{n}\big)+10\big(\vec{S}_1\cdot \vec{n}\big)^2 \big(\vec{S}_2\cdot \vec{v}_2\big)^2
\nonumber\\
&  
+10\big(\vec{S}_1\cdot \vec{n}\big)\big(\vec{S}_1\cdot \vec{v}_1\big) S_2^2 \big(\vec{v}_1\cdot \vec{n}\big) -40 \big(\vec{S}_1\cdot \vec{n}\big) \big(\vec{S}_1\cdot \vec{v}_2\big) S_2^2\big(\vec{v}_2\cdot \vec{n}\big) \nonumber\\
& -70 \big(\vec{S}_1\cdot \vec{n}\big) \big(\vec{S}_1\cdot \vec{v}_1\big) \big(\vec{S}_2\cdot \vec{n}\big)^2 \big(\vec{v}_1\cdot \vec{n}\big) 
+20 \big(\vec{S}_1\cdot \vec{n}\big) \big(\vec{S}_1\cdot \vec{v}_1\big) \big(\vec{S}_2\cdot \vec{n}\big)  \big(\vec{S}_2\cdot
\vec{v}_1\big) \nonumber\\
& +20\big(\vec{S}_1\cdot \vec{n}\big) \big(\vec{S}_1\cdot
\vec{v}_2\big) \big(\vec{S}_2\cdot \vec{n}\big)  \big(\vec{S}_2\cdot \vec{v}_2\big)+20 \big(\vec{S}_1\cdot \vec{n}\big) \big(\vec{S}_1\cdot \vec{S}_2\big)
\big(\vec{S}_2\cdot \vec{v}_2\big) \big(\vec{v}_2\cdot \vec{n}\big) \nonumber\\
& +10\big(\vec{S}_1\cdot \vec{v}_1\big)^2
\big(\vec{S}_2\cdot \vec{n}\big)^2-2\big(\vec{S}_1\cdot \vec{v}_1\big)^2S_2^2 
+20 \big(\vec{S}_1\cdot \vec{v}_1\big)\big(\vec{S}_1\cdot \vec{S}_2\big)\big(\vec{S}_2\cdot \vec{n}\big)
\big(\vec{v}_1\cdot \vec{n}\big)\nonumber\\
&-4 \big(\vec{S}_1\cdot \vec{v}_1\big)\big(\vec{S}_1\cdot
\vec{S}_2\big)  \big(\vec{S}_2\cdot \vec{v}_1\big) -4 \big(\vec{S}_1\cdot \vec{v}_2\big) \big(\vec{S}_1\cdot \vec{S}_2\big)
\big(\vec{S}_2\cdot \vec{v}_2\big)\nonumber\\
& +4 \big(\vec{S}_1\cdot \vec{v}_2\big)^2 S_2^2+4 \big(\vec{S}_1\cdot \vec{S}_2\big)^2\Big(1+\frac{3v_1^2}{2}+\frac{3v_2^2}{2}\Big)\nonumber\\
& -10 S_1^2\big(\vec{S}_2\cdot \vec{n}\big)^2\Big(1+\frac{5v_1^2}{2}+\frac{3v_2^2}{2}-7\big(\vec{v}_1\cdot \vec{n}\big)^2\Big) +4 S_1^2\big(\vec{S}_2\cdot
\vec{v}_1\big)^2 -2  S_1^2\big(\vec{S}_2\cdot \vec{v}_2\big)^2\nonumber\\
& -40  S_1^2\big(\vec{S}_2\cdot \vec{n}\big)
\big(\vec{v}_1\cdot \vec{n}\big) \big(\vec{S}_2\cdot \vec{v}_1\big) +10 S_1^2\big(\vec{S}_2\cdot \vec{n}\big)
 \big(\vec{S}_2\cdot \vec{v}_2\big)\big(\vec{v}_2\cdot \vec{n}\big) \nonumber\\
& +2S_1^2S_2^2\Big(1+\frac{5v_1^2}{2}+\frac{5v_2^2}{2}-5\big(\vec{v}_1\cdot \vec{n}\big)^2-5\big(\vec{v}_2\cdot \vec{n}\big)^2\Big)\Big]\nonumber\\
& +\frac{3C_{1(\text{ES}^2)}C_{2(\text{ES}^2)}G}{2m_1m_2r^4}\Big[5\big(\vec{S}_1\cdot \vec{n}\big)^2 \big(\vec{S}_2\cdot \vec{n}\big)\big(\dot{\vec{S}}_2\cdot
\vec{v}_2\big)\nonumber\\
& -10\big(\vec{S}_1\cdot \vec{n}\big)^2
\big(\dot{\vec{S}}_2\cdot \vec{S}_2\big)  \big(\vec{v}_2\cdot \vec{n}\big)+5 \big(\vec{S}_1\cdot \vec{n}\big)^2 \big(\dot{\vec{S}}_2\cdot\vec{n}\big)  \big(\vec{S}_2\cdot \vec{v}_2\big)\nonumber\\
& +5 \big(\vec{S}_1\cdot \vec{n}\big)^2 \big(\vec{S}_2\cdot\vec{a}_2\big) \big(\vec{S}_2\cdot \vec{n}\big) +\big(\vec{S}_1\cdot\vec{n}\big)\big(\dot{\vec{S}}_1\cdot \vec{v}_1\big)S_2^2 -5 \big(\vec{S}_1\cdot \vec{n}\big)\big(\dot{\vec{S}}_1\cdot \vec{v}_1\big)
\big(\vec{S}_2\cdot \vec{n}\big)^2 \nonumber\\
& -2 \big(\vec{S}_1\cdot \vec{n}\big)\big(\vec{S}_1\cdot\dot{\vec{S}}_2\big)  \big(\vec{S}_2\cdot \vec{v}_2\big) +4\big(\vec{S}_1\cdot \vec{n}\big) \big(\vec{S}_1\cdot \vec{v}_2\big)
\big(\dot{\vec{S}}_2\cdot \vec{S}_2\big) \nonumber\\
& -2 \big(\vec{S}_1\cdot \vec{n}\big) \big(\vec{S}_1\cdot \vec{S}_2\big)\big(\dot{\vec{S}}_2\cdot \vec{v}_2\big)-5 \big(\vec{S}_1\cdot \vec{n}\big) \big(\vec{S}_1\cdot \vec{a}_1\big)  \big(\vec{S}_2\cdot\vec{n}\big)^2 +\big(\vec{S}_1\cdot \vec{n}\big) \big(\vec{S}_1\cdot \vec{a}_1\big) S_2^2\nonumber\\
& -2 \big(\vec{S}_1\cdot\vec{n}\big) \big(\vec{S}_1\cdot \vec{S}_2\big) \big(\vec{S}_2\cdot\vec{a}_2\big) -S_1^2\big(\vec{S}_2\cdot\vec{a}_2\big) \big(\vec{S}_2\cdot \vec{n}\big)+2S_1^2
\big(\dot{\vec{S}}_2\cdot \vec{S}_2\big) \big(\vec{v}_2\cdot \vec{n}\big)   \nonumber\\
& 
+2 \big(\vec{S}_1\cdot \vec{v}_1\big) \big(\dot{\vec{S}}_1\cdot
\vec{S}_2\big) \big(\vec{S}_2\cdot \vec{n}\big)-5 \big(\vec{S}_1\cdot
\vec{v}_1\big) \big(\dot{\vec{S}}_1\cdot \vec{n}\big) \big(\vec{S}_2\cdot \vec{n}\big)^2 + \big(\vec{S}_1\cdot \vec{v}_1\big)\big(\dot{\vec{S}}_1\cdot \vec{n}\big) S_2^2 \nonumber\\
& +2 \big(\vec{S}_1\cdot\vec{a}_1\big) \big(\vec{S}_1\cdot \vec{S}_2\big) \big(\vec{S}_2\cdot \vec{n}\big)
-S_1^2 \big(\vec{S}_2\cdot\vec{n}\big) \big(\dot{\vec{S}}_2\cdot \vec{v}_2\big)-S_1^2\big(\dot{\vec{S}}_2\cdot\vec{n}\big) \big(\vec{S}_2\cdot \vec{v}_2\big)\nonumber\\
&+2 \big(\dot{\vec{S}}_1\cdot \vec{v}_1\big) \big(\vec{S}_2\cdot \vec{n}\big) \big(\vec{S}_1\cdot
\vec{S}_2\big)+10 \big(\dot{\vec{S}}_1\cdot \vec{S}_1\big) \big(\vec{S}_2\cdot \vec{n}\big)^2 \big(\vec{v}_1\cdot \vec{n}\big)\nonumber\\
& -4
\big(\dot{\vec{S}}_1\cdot \vec{S}_1\big) \big(\vec{S}_2\cdot \vec{n}\big) \big(\vec{S}_2\cdot \vec{v}_1\big) -2 \big(\dot{\vec{S}}_1\cdot \vec{S}_1\big) S_2^2
\big(\vec{v}_1\cdot \vec{n}\big) 
\Big]\nonumber\\
& -\frac{C_{1(\text{ES}^2)}C_{2(\text{ES}^2)}G}{m_1m_2r^3}\Big[3 \big(\vec{S}_1\cdot \vec{n}\big)^2\dot{S}_2^2+3 \big(\vec{S}_1\cdot\vec{n}\big)^2 \big(\ddot{\vec{S}}_2\cdot \vec{S}_2\big)  -S_1^2\dot{S}_2^2- S_1^2 \big(\ddot{\vec{S}}_2\cdot \vec{S}_2\big)\nonumber\\
&+3 \dot{S}_1^2 \big(\vec{S}_2\cdot \vec{n}\big)^2 -\dot{S}_1^2 S_2^2 -
\big(\ddot{\vec{S}}_1\cdot \vec{S}_1\big)
S_2^2+3 \big(\ddot{\vec{S}}_1\cdot \vec{S}_1\big) \big(\vec{S}_2\cdot \vec{n}\big)^2
\Big] ,\\
\text{Fig.~2(a2)}&=\frac{3C_{1(\text{ES}^2)}C_{2(\text{ES}^2)}G}{8m_1m_2r^5}\Big[5 \big(\vec{S}_1\cdot \vec{n}\big)^2 S_2^2 
\big(\vec{v}_1\cdot \vec{v}_2\big)
-35 \big(\vec{S}_1\cdot \vec{n}\big)^2 S_2^2 \big(\vec{v}_1\cdot \vec{n}\big) \big(\vec{v}_2\cdot \vec{n}\big)\nonumber\\
&\quad -315 \big(\vec{S}_1\cdot \vec{n}\big)^2\big(\vec{S}_2\cdot \vec{n}\big)^2
\big(\vec{v}_1\cdot \vec{n}\big) \big(\vec{v}_2\cdot \vec{n}\big) +70 \big(\vec{S}_1\cdot \vec{n}\big)^2\big(\vec{S}_2\cdot\vec{n}\big) \big(\vec{S}_2\cdot \vec{v}_1\big)\big(\vec{v}_2\cdot \vec{n}\big) \nonumber\\
&\quad  +70 \big(\vec{S}_1\cdot\vec{n}\big)^2\big(\vec{S}_2\cdot \vec{n}\big)\big(\vec{S}_2\cdot \vec{v}_2\big) \big(\vec{v}_1\cdot \vec{n}\big)  -10 \big(\vec{S}_1\cdot\vec{n}\big)^2 \big(\vec{S}_2\cdot \vec{v}_1\big) \big(\vec{S}_2\cdot \vec{v}_2\big)\nonumber\\
&\quad +35 \big(\vec{S}_1\cdot\vec{n}\big)^2\big(\vec{S}_2\cdot \vec{n}\big)^2 \big(\vec{v}_1\cdot \vec{v}_2\big) +10\big(\vec{S}_1\cdot\vec{n}\big)  \big(\vec{S}_1\cdot \vec{v}_1\big) S_2^2 \big(\vec{v}_2\cdot \vec{n}\big)\nonumber\\
&\quad +10 \big(\vec{S}_1\cdot\vec{n}\big) \big(\vec{S}_1\cdot \vec{v}_2\big) S_2^2  \big(\vec{v}_1\cdot \vec{n}\big)+140 \big(\vec{S}_1\cdot \vec{n}\big)\big(\vec{S}_1\cdot \vec{S}_2\big)\big(\vec{S}_2\cdot \vec{n}\big) \big(\vec{v}_1\cdot \vec{n}\big) \big(\vec{v}_2\cdot \vec{n}\big)
 \nonumber\\
&\quad +70 \big(\vec{S}_1\cdot \vec{n}\big)\big(\vec{S}_1\cdot \vec{v}_1\big)\big(\vec{S}_2\cdot \vec{n}\big)^2 \big(\vec{v}_2\cdot\vec{n}\big)  +70 \big(\vec{S}_1\cdot \vec{n}\big)\big(\vec{S}_1\cdot \vec{v}_2\big) \big(\vec{S}_2\cdot \vec{n}\big)^2
\big(\vec{v}_1\cdot \vec{n}\big) \nonumber\\
&\quad -20 \big(\vec{S}_1\cdot \vec{n}\big) \big(\vec{S}_1\cdot \vec{S}_2\big) \big(\vec{S}_2\cdot \vec{v}_1\big)\big(\vec{v}_2\cdot\vec{n}\big) -20 \big(\vec{S}_1\cdot\vec{n}\big)\big(\vec{S}_1\cdot \vec{v}_2\big)
\big(\vec{S}_2\cdot \vec{n}\big) \big(\vec{S}_2\cdot \vec{v}_1\big) \nonumber\\
&\quad -20 \big(\vec{S}_1\cdot \vec{n}\big) \big(\vec{S}_1\cdot \vec{S}_2\big) \big(\vec{S}_2\cdot \vec{v}_2\big)\big(\vec{v}_1\cdot \vec{n}\big)
 -20 \big(\vec{S}_1\cdot \vec{n}\big)  \big(\vec{S}_1\cdot \vec{v}_1\big)\big(\vec{S}_2\cdot \vec{n}\big) \big(\vec{S}_2\cdot
\vec{v}_2\big)\nonumber\\
&\quad -20 \big(\vec{S}_1\cdot \vec{n}\big) \big(\vec{S}_1\cdot \vec{S}_2\big) \big(\vec{S}_2\cdot \vec{n}\big) 
\big(\vec{v}_1\cdot \vec{v}_2\big)+4\big(\vec{S}_1\cdot \vec{v}_1\big)  \big(\vec{S}_1\cdot
\vec{S}_2\big) \big(\vec{S}_2\cdot \vec{v}_2\big)\nonumber\\
&\quad -20 \big(\vec{S}_1\cdot \vec{v}_1\big) \big(\vec{S}_1\cdot \vec{S}_2\big) \big(\vec{S}_2\cdot\vec{n}\big) \big(\vec{v}_2\cdot \vec{n}\big)-10 \big(\vec{S}_1\cdot \vec{v}_1\big) \big(\vec{S}_1\cdot \vec{v}_2\big)\big(\vec{S}_2\cdot \vec{n}\big)^2\nonumber\\
&\quad -2 \big(\vec{S}_1\cdot \vec{v}_1\big) \big(\vec{S}_1\cdot \vec{v}_2\big)S_2^2 -20\big(\vec{S}_1\cdot
\vec{v}_2\big)\big(\vec{S}_1\cdot \vec{S}_2\big)
\big(\vec{S}_2\cdot \vec{n}\big) \big(\vec{v}_1\cdot \vec{n}\big)\nonumber\\  &\quad +4\big(\vec{S}_1\cdot \vec{v}_2\big)
\big(\vec{S}_1\cdot \vec{S}_2\big)  \big(\vec{S}_2\cdot \vec{v}_1\big) -10 \big(\vec{S}_1\cdot \vec{S}_2\big)^2 \big(\vec{v}_1\cdot \vec{n}\big) \big(\vec{v}_2\cdot\vec{n}\big)\nonumber\\
&\quad +2 \big(\vec{S}_1\cdot \vec{S}_2\big)^2
\big(\vec{v}_1\cdot \vec{v}_2\big) -35 S_1^2 \big(\vec{S}_2\cdot \vec{n}\big)^2 \big(\vec{v}_1\cdot \vec{n}\big)
\big(\vec{v}_2\cdot \vec{n}\big)+5 S_1^2\big(\vec{S}_2\cdot \vec{n}\big)^2 \big(\vec{v}_1\cdot \vec{v}_2\big)\nonumber\\
&\quad +10 S_1^2\big(\vec{S}_2\cdot \vec{n}\big)\big(\vec{S}_2\cdot \vec{v}_1\big) \big(\vec{v}_2\cdot \vec{n}\big) +10 S_1^2\big(\vec{S}_2\cdot \vec{n}\big) \big(\vec{v}_1\cdot \vec{n}\big)
\big(\vec{S}_2\cdot \vec{v}_2\big)\nonumber\\
&\quad -2 S_1^2\big(\vec{S}_2\cdot \vec{v}_1\big) \big(\vec{S}_2\cdot \vec{v}_2\big) +15 S_1^2 S_2^2\big(\vec{v}_1\cdot\vec{n}\big) \big(\vec{v}_2\cdot \vec{n}\big) -3 S_1^2 S_2^2\big(\vec{v}_1\cdot \vec{v}_2\big)
\Big]\nonumber\\
&\quad +\frac{3C_{1(\text{ES}^2)}C_{2(\text{ES}^2)}G}{4m_1m_2r^4}\Big[5\big(\vec{S}_1\cdot\vec{n}\big)^2 \big(\vec{S}_2\cdot \vec{n}\big) \big(\dot{\vec{S}}_2\cdot \vec{v}_1\big) \nonumber\\
&\quad 
-35 \big(\vec{S}_1\cdot\vec{n}\big)^2\big(\vec{S}_2\cdot \vec{n}\big)\big(\dot{\vec{S}}_2\cdot \vec{n}\big)  \big(\vec{v}_1\cdot \vec{n}\big)  -5 \big(\vec{S}_1\cdot\vec{n}\big)^2 \big(\dot{\vec{S}}_2\cdot \vec{S}_2\big)\big(\vec{v}_1\cdot \vec{n}\big) \nonumber\\
&\quad +5 \big(\vec{S}_1\cdot\vec{n}\big)^2\big(\vec{S}_2\cdot \vec{v}_1\big)\big(\dot{\vec{S}}_2\cdot \vec{n}\big)  +5\big(\vec{S}_1\cdot\vec{n}\big) \big(\dot{\vec{S}}_1\cdot \vec{n}\big)S_2^2 \big(\vec{v}_2\cdot \vec{n}\big)  - \big(\vec{S}_1\cdot \vec{n}\big)\big(\dot{\vec{S}}_1\cdot \vec{v}_2\big) S_2^2\nonumber\\
&\quad +35 \big(\vec{S}_1\cdot \vec{n}\big)\big(\dot{\vec{S}}_1\cdot\vec{n}\big) \big(\vec{S}_2\cdot \vec{n}\big){}^2 \big(\vec{v}_2\cdot \vec{n}\big) -10\big(\vec{S}_1\cdot\vec{n}\big)\big(\dot{\vec{S}}_1\cdot \vec{S}_2\big)
\big(\vec{S}_2\cdot \vec{n}\big) \big(\vec{v}_2\cdot \vec{n}\big)  \nonumber\\
&\quad -5 \big(\vec{S}_1\cdot\vec{n}\big)\big(\dot{\vec{S}}_1\cdot \vec{v}_2\big)\big(\vec{S}_2\cdot \vec{n}\big){}^2  +10 \big(\vec{S}_1\cdot \vec{n}\big)\big(\vec{S}_1\cdot\dot{\vec{S}}_2\big)\big(\vec{S}_2\cdot \vec{n}\big) \big(\vec{v}_1\cdot \vec{n}\big) 
\nonumber\\
&\quad +10 \big(\vec{S}_1\cdot \vec{n}\big) \big(\vec{S}_1\cdot
\vec{S}_2\big) \big(\dot{\vec{S}}_2\cdot \vec{n}\big) \big(\vec{v}_1\cdot \vec{n}\big)-2 \big(\vec{S}_1\cdot \vec{n}\big) \big(\vec{S}_1\cdot \vec{S}_2\big)
\big(\dot{\vec{S}}_2\cdot \vec{v}_1\big)\nonumber\\
&\quad +10 \big(\vec{S}_1\cdot \vec{n}\big) \big(\vec{S}_1\cdot
\vec{v}_1\big)\big(\vec{S}_2\cdot \vec{n}\big) \big(\dot{\vec{S}}_2\cdot \vec{n}\big)  +2 \big(\vec{S}_1\cdot \vec{n}\big)\big(\vec{S}_1\cdot \vec{v}_1\big)\big(\dot{\vec{S}}_2\cdot \vec{S}_2\big) 
\nonumber\\
&\quad -2\big(\vec{S}_1\cdot\vec{n}\big) \big(\vec{S}_1\cdot\dot{\vec{S}}_2\big) \big(\vec{S}_2\cdot \vec{v}_1\big) -10\big(\vec{S}_1\cdot \vec{n}\big) \big(\dot{\vec{S}}_1\cdot \vec{n}\big) \big(\vec{S}_2\cdot \vec{n}\big) \big(\vec{S}_2\cdot \vec{v}_2\big)
\nonumber\\
&\quad +2 \big(\vec{S}_1\cdot\vec{n}\big)\big(\dot{\vec{S}}_1\cdot \vec{S}_2\big) \big(\vec{S}_2\cdot \vec{v}_2\big) -2\big(\vec{S}_1\cdot \vec{v}_1\big) \big(\vec{S}_1\cdot\dot{\vec{S}}_2\big)\big(\vec{S}_2\cdot \vec{n}\big)\nonumber\\
&\quad -2\big(\vec{S}_1\cdot \vec{v}_1\big) \big(\vec{S}_1\cdot \vec{S}_2\big)\big(\dot{\vec{S}}_2\cdot \vec{n}\big)-5 \big(\vec{S}_1\cdot \vec{v}_2\big)\big(\dot{\vec{S}}_1\cdot \vec{n}\big) \big(\vec{S}_2\cdot\vec{n}\big)^2\nonumber\\
&\quad +2 \big(\vec{S}_1\cdot \vec{v}_2\big)\big(\dot{\vec{S}}_1\cdot
\vec{S}_2\big)\big(\vec{S}_2\cdot \vec{n}\big) -2\big(\vec{S}_2\cdot \vec{v}_2\big)\big(\dot{\vec{S}}_1\cdot \vec{S}_1\big) \big(\vec{S}_2\cdot \vec{n}\big) \nonumber\\
&\quad +2 \big(\vec{S}_2\cdot \vec{v}_2\big)\big(\vec{S}_1\cdot \vec{S}_2\big)\big(\dot{\vec{S}}_1\cdot \vec{n}\big)-\big(\vec{S}_1\cdot \vec{v}_2\big)\big(\dot{\vec{S}}_1\cdot \vec{n}\big) S_2^2\nonumber\\
&\quad -10 \big(\vec{S}_1\cdot \vec{S}_2\big) \big(\dot{\vec{S}}_1\cdot\vec{n}\big) \big(\vec{S}_2\cdot \vec{n}\big) \big(\vec{v}_2\cdot \vec{n}\big) +2\big(\vec{S}_1\cdot \vec{S}_2\big)
\big(\dot{\vec{S}}_1\cdot \vec{S}_2\big)\big(\vec{v}_2\cdot \vec{n}\big)\nonumber\\
&\quad +2 \big(\vec{S}_1\cdot \vec{S}_2\big) \big(\dot{\vec{S}}_1\cdot \vec{v}_2\big) \big(\vec{S}_2\cdot\vec{n}\big) -2\big(\vec{S}_1\cdot \vec{S}_2\big) 
\big(\vec{S}_1\cdot\dot{\vec{S}}_2\big)\big(\vec{v}_1\cdot \vec{n}\big)\nonumber\\
&\quad -5 S_1^2\big(\dot{\vec{S}}_2\cdot \vec{n}\big) \big(\vec{S}_2\cdot \vec{n}\big) \big(\vec{v}_1\cdot \vec{n}\big) +3 S_1^2\big(\dot{\vec{S}}_2\cdot \vec{S}_2\big)\big(\vec{v}_1\cdot \vec{n}\big)  +S_1^2\big(\vec{S}_2\cdot\vec{n}\big) \big(\dot{\vec{S}}_2\cdot \vec{v}_1\big)\nonumber\\
&\quad +S_1^2\big(\dot{\vec{S}}_2\cdot \vec{n}\big) \big(\vec{S}_2\cdot
\vec{v}_1\big) -3 \big(\dot{\vec{S}}_1\cdot \vec{S}_1\big)S_2^2 \big(\vec{v}_2\cdot \vec{n}\big)
 +5 \big(\dot{\vec{S}}_1\cdot \vec{S}_1\big)\big(\vec{S}_2\cdot\vec{n}\big)^2 \big(\vec{v}_2\cdot \vec{n}\big) 
\Big]\nonumber\\
&\quad +\frac{C_{1(\text{ES}^2)}C_{2(\text{ES}^2)}G}{2m_1m_2r^3}\Big[3 \big(\vec{S}_1\cdot\vec{n}\big)\big(\dot{\vec{S}}_1\cdot \vec{n}\big)\big(\dot{\vec{S}}_2\cdot \vec{S}_2\big)-3\big(\vec{S}_1\cdot \vec{n}\big) \big(\dot{\vec{S}}_1\cdot \dot{\vec{S}}_2\big)  \big(\vec{S}_2\cdot\vec{n}\big)\nonumber\\
&\quad -3 \big(\vec{S}_1\cdot\vec{n}\big)\big(\dot{\vec{S}}_1\cdot \vec{S}_2\big) \big(\dot{\vec{S}}_2\cdot \vec{n}\big) +15\big(\vec{S}_1\cdot \vec{n}\big) \big(\dot{\vec{S}}_1\cdot \vec{n}\big) \big(\vec{S}_2\cdot \vec{n}\big)\big(\dot{\vec{S}}_2\cdot \vec{n}\big) \nonumber\\
&\quad +\big(\vec{S}_1\cdot \vec{S}_2\big)\big(\dot{\vec{S}}_1\cdot \dot{\vec{S}}_2\big) -3 \big(\vec{S}_1\cdot \vec{S}_2\big)\big(\dot{\vec{S}}_1\cdot \vec{n}\big) \big(\dot{\vec{S}}_2\cdot \vec{n}\big)+3 \big(\dot{\vec{S}}_1\cdot \vec{S}_1\big) \big(\vec{S}_2\cdot \vec{n}\big)\big(\dot{\vec{S}}_2\cdot\vec{n}\big) \nonumber\\
&\quad  -3 \big(\dot{\vec{S}}_1\cdot
\vec{S}_1\big) \big(\dot{\vec{S}}_2\cdot \vec{S}_2\big)+ \big(\vec{S}_1\cdot\dot{\vec{S}}_2\big)\big(\dot{\vec{S}}_1\cdot \vec{S}_2\big) -3 \big(\vec{S}_1\cdot\dot{\vec{S}}_2\big) \big(\dot{\vec{S}}_1\cdot \vec{n}\big) \big(\vec{S}_2\cdot\vec{n}\big)
\Big] ,&\\
\text{Fig.~2(a3)}&=\frac{3C_{1(\text{ES}^2)}C_{2(\text{ES}^2)}G}{m_1m_2r^5}\Big[
5 
\big(\vec{S}_1\cdot \vec{n}\big)^2 S_2^2 \big(\vec{v}_1\cdot \vec{v}_2\big)-35 \big(\vec{S}_1\cdot \vec{n}\big){}^2 \big(\vec{S}_2\cdot \vec{n}\big)^2\nonumber\\
&\quad +20 \big(\vec{S}_1\cdot \vec{n}\big) \big(\vec{S}_1\cdot \vec{S}_2\big)\big(\vec{S}_2\cdot \vec{n}\big) -S_1^2S_2^2 +5 S_1^2\big(\vec{S}_2\cdot\vec{n}\big)^2 -2 \big(\vec{S}_1\cdot \vec{S}_2\big)^2
\Big]\big(\vec{v}_1\cdot
\vec{v}_2\big)\nonumber\\
&\quad +\frac{3C_{1(\text{ES}^2)}C_{2(\text{ES}^2)}G}{m_1m_2r^4} \Big[ 10\big(\vec{S}_1\cdot \vec{n}\big)^2
\big(\dot{\vec{S}}_2\cdot \vec{S}_2\big)  \big(\vec{v}_1\cdot \vec{n}\big) -5 \big(\vec{S}_1\cdot \vec{n}\big)^2\big(\vec{S}_2\cdot \vec{v}_1\big) \big(\dot{\vec{S}}_2\cdot\vec{n}\big)\nonumber\\
&\quad -5\big(\vec{S}_1\cdot \vec{n}\big)^2  \big(\vec{S}_2\cdot \vec{n}\big)\big(\dot{\vec{S}}_2\cdot
\vec{v}_1\big)-\big(\vec{S}_1\cdot\vec{n}\big)\big(\dot{\vec{S}}_1\cdot \vec{v}_2\big)  S_2^2 -4\big(\vec{S}_1\cdot \vec{n}\big)\big(\vec{S}_1\cdot \vec{v}_1\big)
\big(\dot{\vec{S}}_2\cdot \vec{S}_2\big)\nonumber\\
&\quad+5 \big(\vec{S}_1\cdot \vec{n}\big)\big(\dot{\vec{S}}_1\cdot \vec{v}_2\big)
\big(\vec{S}_2\cdot \vec{n}\big)^2+2 
\big(\vec{S}_1\cdot \vec{n}\big) \big(\vec{S}_1\cdot \vec{S}_2\big)\big(\dot{\vec{S}}_2\cdot \vec{v}_1\big)\nonumber\\
&\quad+2 \big(\vec{S}_1\cdot \vec{n}\big)\big(\vec{S}_1\cdot\dot{\vec{S}}_2\big)  \big(\vec{S}_2\cdot \vec{v}_1\big)-2 \big(\vec{S}_1\cdot \vec{v}_2\big) \big(\dot{\vec{S}}_1\cdot
\vec{S}_2\big) \big(\vec{S}_2\cdot \vec{n}\big)-\big(\vec{S}_1\cdot \vec{v}_2\big)\big(\dot{\vec{S}}_1\cdot \vec{n}\big) S_2^2\nonumber\\
&\quad +5\big(\vec{S}_1\cdot
\vec{v}_2\big) \big(\dot{\vec{S}}_1\cdot \vec{n}\big) \big(\vec{S}_2\cdot \vec{n}\big)^2-2 \big(\vec{S}_1\cdot
\vec{S}_2\big) \big(\dot{\vec{S}}_1\cdot \vec{v}_2\big) \big(\vec{S}_2\cdot \vec{n}\big) \nonumber\\
&\quad +S_1^2\big(\vec{S}_2\cdot \vec{v}_1\big) \big(
\dot{\vec{S}}_2\cdot\vec{n}\big) -2S_1^2
\big(\dot{\vec{S}}_2\cdot \vec{S}_2\big) \big(\vec{v}_1\cdot \vec{n}\big) + S_1^2\big(\vec{S}_2\cdot\vec{n}\big)\big(\dot{\vec{S}}_2\cdot \vec{v}_1\big)\nonumber\\
&\quad +4
\big(\dot{\vec{S}}_1\cdot \vec{S}_1\big) \big(\vec{S}_2\cdot \vec{n}\big) \big(\vec{S}_2\cdot \vec{v}_2\big) -10 \big(\dot{\vec{S}}_1\cdot \vec{S}_1\big) \big(\vec{S}_2\cdot \vec{n}\big)^2 \big(\vec{v}_2\cdot \vec{n}\big)\nonumber\\
&\quad +2 
\big(\dot{\vec{S}}_1\cdot \vec{S}_1\big) S_2^2\big(\vec{v}_2\cdot \vec{n}\big)
\Big] +\frac{C_{1(\text{ES}^2)}C_{2(\text{ES}^2)}G}{m_1m_2r^3}\Big[
3 \big(\vec{S}_1\cdot \vec{n}\big)\big(\dot{\vec{S}}_1\cdot \dot{\vec{S}}_2\big)  \big(\vec{S}_2\cdot \vec{n}\big)\nonumber\\
&\quad -12\big(\vec{S}_1\cdot \vec{n}\big)\big(\dot{\vec{S}}_1\cdot \vec{n}\big)
\big(\dot{\vec{S}}_2\cdot \vec{S}_2\big) +3\big(\vec{S}_1\cdot \vec{n}\big)
\big(\dot{\vec{S}}_1\cdot \vec{S}_2\big) \big(\dot{\vec{S}}_2\cdot \vec{n}\big)\nonumber\\
&\quad +3\big(\vec{S}_1\cdot \vec{S}_2\big)
\big(\dot{\vec{S}}_1\cdot \vec{n}\big) \big(\dot{\vec{S}}_2\cdot \vec{n}\big)  -2\big(\vec{S}_1\cdot \vec{S}_2\big)
\big(\dot{\vec{S}}_1\cdot \dot{\vec{S}}_2\big) +8 \big(\dot{\vec{S}}_1\cdot \vec{S}_1\big)
\big(\dot{\vec{S}}_2\cdot \vec{S}_2\big)\nonumber\\
&\quad -12 \big(\dot{\vec{S}}_1\cdot \vec{S}_1\big) \big(\dot{\vec{S}}_2\cdot \vec{n}\big)
\big(\vec{S}_2\cdot \vec{n}\big) -2 \big(\dot{\vec{S}}_1\cdot \vec{S}_2\big) \big(\dot{\vec{S}}_2\cdot \vec{S}_1\big) \nonumber\\*
&\quad +3
\big(\vec{S}_1\cdot\dot{\vec{S}}_2\big) \big(\dot{\vec{S}}_1\cdot \vec{n}\big) \big(\vec{S}_2\cdot \vec{n}\big)
\Big] ,&\\
\text{Fig.~2(b1)}&=\frac{C_{1(\text{BS}^3)}}{4}\frac{G}{m_1^2r^5}\Big[
70 \big(\vec{S}_1\cdot \vec{n}\big)^3\big(\vec{S}_2\cdot \vec{n}\big)\Big(1+\frac{3v_1^2}{2}+\frac{3v_2^2}{2}\Big) \nonumber\\*
&\quad -30 \big(\vec{S}_1\cdot \vec{n}\big)^2\big(\vec{S}_1\cdot \vec{S}_2\big)\Big(1+\frac{13v_1^2}{6}+\frac{13v_2^2}{6}-\frac{14\big(\vec{v}_1\cdot \vec{n}\big)^2}{3}-\frac{14\big(\vec{v}_2\cdot \vec{n}\big)^2}{3}\Big)\nonumber\\*
&\quad +6 S_1^2 \big(\vec{S}_1\cdot \vec{S}_2\big)\Big(1+\frac{13v_1^2}{6}+\frac{13v_2^2}{6}-\frac{10\big(\vec{v}_1\cdot \vec{n}\big){}^2}{3}-\frac{10\big(\vec{v}_2\cdot \vec{n}\big){}^2}{3}\Big) \nonumber\\
&\quad -30 S_1^2 \big(\vec{S}_1\cdot \vec{n}\big)
\big(\vec{S}_2\cdot \vec{n}\big)\Big(1+\frac{3v_1^2}{2}+\frac{3v_2^2}{2}\Big)-2 \big(\vec{S}_1\cdot \vec{v}_1\big)^2 \big(\vec{S}_1\cdot \vec{S}_2\big) \nonumber\\*
&\quad -140 \big(\vec{S}_1\cdot \vec{n}\big)^3 \big(\vec{S}_2\cdot
\vec{v}_1\big) \big(\vec{v}_1\cdot \vec{n}\big)-105 \big(\vec{S}_1\cdot \vec{n}\big)^3\big(\vec{S}_2\cdot \vec{v}_2\big)
\big(\vec{v}_2\cdot \vec{n}\big) \nonumber\\*
&\quad -140  \big(\vec{S}_1\cdot\vec{n}\big)^2\big(\vec{S}_1\cdot \vec{v}_2\big)
\big(\vec{S}_2\cdot \vec{n}\big) \big(\vec{v}_2\cdot \vec{n}\big)+85 \big(\vec{S}_1\cdot\vec{n}\big)^2 \big(\vec{S}_1\cdot \vec{v}_1\big) \big(\vec{S}_2\cdot \vec{v}_1\big) \nonumber\\
&\quad -175 \big(\vec{S}_1\cdot \vec{n}\big)^2\big(\vec{S}_1\cdot \vec{v}_1\big)\big(\vec{S}_2\cdot\vec{n}\big) \big(\vec{v}_1\cdot \vec{n}\big)  +65 \big(\vec{S}_1\cdot\vec{n}\big)^2 \big(\vec{S}_1\cdot \vec{v}_2\big) \big(\vec{S}_2\cdot \vec{v}_2\big) \nonumber\\
&\quad  +50 \big(\vec{S}_1\cdot\vec{n}\big)\big(\vec{S}_1\cdot \vec{v}_1\big){}^2 \big(\vec{S}_2\cdot \vec{n}\big) +40 \big(\vec{S}_1\cdot \vec{n}\big) \big(\vec{S}_1\cdot \vec{v}_2\big)^2 \big(\vec{S}_2\cdot \vec{n}\big) \nonumber\\
&\quad  -30 \big(\vec{S}_1\cdot \vec{n}\big)\big(\vec{S}_1\cdot \vec{v}_1\big)\big(\vec{S}_1\cdot \vec{S}_2\big)\big(\vec{v}_1\cdot\vec{n}\big)-40 \big(\vec{S}_1\cdot\vec{n}\big)\big(\vec{S}_1\cdot \vec{v}_2\big) \big(\vec{S}_1\cdot \vec{S}_2\big)\big(\vec{v}_2\cdot \vec{n}\big)\nonumber\\
&\quad +60 S_1^2 \big(\vec{S}_1\cdot \vec{n}\big) \big(\vec{S}_2\cdot \vec{v}_1\big)
\big(\vec{v}_1\cdot \vec{n}\big)+45 S_1^2 \big(\vec{S}_1\cdot \vec{n}\big) \big(\vec{S}_2\cdot
\vec{v}_2\big) \big(\vec{v}_2\cdot \vec{n}\big)\nonumber\\
&\quad +25 S_1^2\big(\vec{S}_2\cdot\vec{n}\big)  \big(\vec{S}_1\cdot \vec{v}_1\big)\big(\vec{v}_1\cdot \vec{n}\big)-13
S_1^2 \big(\vec{S}_1\cdot \vec{v}_2\big) \big(\vec{S}_2\cdot \vec{v}_2\big)\nonumber\\
&\quad +20 S_1^2
\big(\vec{S}_2\cdot \vec{n}\big) \big(\vec{S}_1\cdot
\vec{v}_2\big) \big(\vec{v}_2\cdot \vec{n}\big) -17 S_1^2 \big(\vec{S}_1\cdot \vec{v}_1\big) \big(\vec{S}_2\cdot \vec{v}_1\big)\nonumber\\
&\quad +\frac{C_{1(\text{BS}^3)}}{2}\frac{G}{m_1^2r^4}\Big[
10 \big(\vec{S}_1\cdot \vec{n}\big)^3\big(\vec{S}_2\cdot \vec{a}_1\big) \nonumber\\
&\quad -10 \big(\vec{S}_1\cdot \vec{n}\big)^2\big(\vec{S}_1\cdot\vec{a}_2\big)\big(\vec{S}_2\cdot \vec{n}\big)-10 \big(\vec{S}_1\cdot \vec{n}\big)^2\big(\dot{\vec{S}}_1\cdot \vec{S}_2\big)
\big(\vec{v}_1\cdot \vec{n}\big) \nonumber\\
&\quad +10 \big(\vec{S}_1\cdot \vec{n}\big)^2\big(\vec{S}_1\cdot\dot{\vec{S}}_2\big)
\big(\vec{v}_2\cdot \vec{n}\big)-10 \big(\vec{S}_1\cdot\vec{n}\big)^2 \big(\vec{S}_1\cdot \vec{S}_2\big) \big(\vec{a}_1\cdot\vec{n}\big)\nonumber\\
&\quad +10 \big(\vec{S}_1\cdot\vec{n}\big)^2 \big(\vec{S}_1\cdot \vec{S}_2\big)\big(\vec{a}_2\cdot\vec{n}\big) -10 \big(\vec{S}_1\cdot\vec{n}\big)^2 \big(\vec{S}_1\cdot \vec{v}_2\big) \big(\dot{\vec{S}}_2\cdot\vec{n}\big)  \nonumber\\
&\quad +30 \big(\vec{S}_1\cdot\vec{n}\big)^2 \big(\dot{\vec{S}}_1\cdot\vec{n}\big) \big(\vec{S}_2\cdot \vec{v}_1\big) +10 \big(\vec{S}_1\cdot \vec{n}\big) \big(\dot{\vec{S}}_1\cdot \vec{S}_1\big) \big(\vec{S}_2\cdot \vec{n}\big) \big(\vec{v}_1\cdot \vec{n}\big) 
\nonumber\\
&\quad +5 S_1^2 \big(\vec{S}_1\cdot \vec{n}\big) \big(\vec{S}_2\cdot \vec{n}\big) \big(\vec{a}_1\cdot\vec{n}\big)  -7 S_1^2\big(\vec{S}_1\cdot \vec{n}\big)\big(\vec{S}_2\cdot \vec{a}_1\big) 
\nonumber\\
&\quad -20 \big(\vec{S}_1\cdot \vec{n}\big)\big(\dot{\vec{S}}_1\cdot \vec{n}\big)  \big(\vec{S}_1\cdot
\vec{S}_2\big) \big(\vec{v}_1\cdot \vec{n}\big)+4\big(\vec{S}_1\cdot \vec{n}\big) \big(\vec{S}_1\cdot \vec{a}_1\big) \big(\vec{S}_1\cdot \vec{S}_2\big)
\nonumber\\
&\quad +4 \big(\vec{S}_1\cdot\vec{n}\big)\big(\dot{\vec{S}}_1\cdot \vec{v}_1\big) \big(\vec{S}_1\cdot \vec{S}_2\big) +4 \big(\vec{S}_1\cdot \vec{n}\big)\big(\vec{S}_1\cdot \vec{v}_1\big)\big(\dot{\vec{S}}_1\cdot \vec{S}_2\big)  \nonumber\\
&\quad -14 \big(\vec{S}_1\cdot \vec{n}\big)
\big(\dot{\vec{S}}_1\cdot \vec{S}_1\big) \big(\vec{S}_2\cdot \vec{v}_1\big) +5 S_1^2\big(\dot{\vec{S}}_1\cdot\vec{n}\big) \big(\vec{S}_2\cdot \vec{n}\big) \big(\vec{v}_1\cdot \vec{n}\big)\nonumber\\
&\quad +2 S_1^2 \big(\vec{S}_1\cdot \vec{a}_2\big) 
\big(\vec{S}_2\cdot \vec{n}\big) -S_1^2\big(\vec{S}_1\cdot \vec{a}_1\big)\big(\vec{S}_2\cdot \vec{n}\big) +4 \big(\dot{\vec{S}}_1\cdot \vec{n}\big)\big(\vec{S}_1\cdot \vec{v}_1\big) \big(\vec{S}_1\cdot \vec{S}_2\big)
 \nonumber\\
&\quad -2 S_1^2
\big(\vec{S}_1\cdot\dot{\vec{S}}_2\big) \big(\vec{v}_2\cdot \vec{n}\big)
+2\big(\vec{S}_1\cdot\dot{\vec{S}}_1\big)\big(\vec{S}_1\cdot \vec{S}_2\big)\big(\vec{v}_1\cdot \vec{n}\big) 
+S_1^2  \big(\vec{S}_1\cdot \vec{S}_2\big)\big(\vec{a}_1\cdot\vec{n}\big)\nonumber\\
&\quad -2 S_1^2
\big(\vec{S}_1\cdot \vec{S}_2\big)\big(\vec{a}_2\cdot\vec{n}\big)  -2  \big(\vec{S}_1\cdot \dot{\vec{S}}_1\big)
\big(\vec{S}_1\cdot \vec{v}_1\big)\big(\vec{S}_2\cdot \vec{n}\big)-S_1^2 \big(\dot{\vec{S}}_1\cdot \vec{v}_1\big) \big(\vec{S}_2\cdot \vec{n}\big)
\nonumber\\
&\quad +S_1^2 \big(\dot{\vec{S}}_1\cdot \vec{S}_2\big) \big(\vec{v}_1\cdot\vec{n}\big) +2 S_1^2
\big(\vec{S}_1\cdot \vec{v}_2\big)\big(\dot{\vec{S}}_2\cdot\vec{n}\big) -7S_1^2 \big(\dot{\vec{S}}_1\cdot \vec{n}\big) 
\big(\vec{S}_2\cdot \vec{v}_1\big)
\Big] ,&\\
\text{Fig.~2(b2)}&=\frac{C_{1(\text{BS}^3)}}{4}\frac{G}{m_1^2 r^5}\Big[
35 \big(\vec{S}_1\cdot\vec{n}\big)^3 \big(\vec{S}_2\cdot \vec{v}_1\big)\big(\vec{v}_2\cdot \vec{n}\big) +35 \big(\vec{S}_1\cdot\vec{n}\big)^3 \big(\vec{S}_2\cdot \vec{v}_2\big) \big(\vec{v}_1\cdot \vec{n}\big)\nonumber\\
&\quad +35 \big(\vec{S}_1\cdot\vec{n}\big)^3\big(\vec{S}_2\cdot \vec{n}\big) \big(\vec{v}_1\cdot \vec{v}_2\big) -315 \big(\vec{S}_1\cdot\vec{n}\big)^3 \big(\vec{S}_2\cdot \vec{n}\big) \big(\vec{v}_1\cdot \vec{n}\big) \big(\vec{v}_2\cdot \vec{n}\big) \nonumber\\
&\quad +35 \big(\vec{S}_1\cdot \vec{n}\big)^2\big(\vec{S}_1\cdot \vec{S}_2\big)\big(\vec{v}_1\cdot \vec{n}\big) \big(\vec{v}_2\cdot \vec{n}\big) +105\big(\vec{S}_1\cdot \vec{n}\big)^2 \big(\vec{S}_2\cdot \vec{n}\big) \big(\vec{v}_2\cdot \vec{n}\big) \big(\vec{S}_1\cdot
\vec{v}_1\big) \nonumber\\
&\quad +105 \big(\vec{S}_1\cdot \vec{n}\big)^2\big(\vec{S}_1\cdot \vec{v}_2\big) \big(\vec{S}_2\cdot \vec{n}\big) \big(\vec{v}_1\cdot \vec{n}\big)
-15\big(\vec{S}_1\cdot \vec{n}\big)^2 \big(\vec{S}_1\cdot \vec{v}_2\big)
\big(\vec{S}_2\cdot \vec{v}_1\big) \nonumber\\
&\quad -15 \big(\vec{S}_1\cdot \vec{n}\big)^2\big(\vec{S}_1\cdot \vec{v}_1\big)
\big(\vec{S}_2\cdot \vec{v}_2\big) -5\big(\vec{S}_1\cdot \vec{n}\big)^2 \big(\vec{S}_1\cdot \vec{S}_2\big) \big(\vec{v}_1\cdot
\vec{v}_2\big) \nonumber\\
&\quad -10 \big(\vec{S}_1\cdot \vec{n}\big)\big(\vec{S}_1\cdot \vec{v}_1\big) \big(\vec{S}_1\cdot \vec{S}_2\big)  \big(\vec{v}_2\cdot\vec{n}\big)-30\big(\vec{S}_1\cdot \vec{n}\big)  \big(\vec{S}_1\cdot \vec{v}_1\big) \big(\vec{S}_1\cdot \vec{v}_2\big)\big(\vec{S}_2\cdot \vec{n}\big)  \nonumber\\
&\quad -10\big(\vec{S}_1\cdot\vec{n}\big)\big(\vec{S}_1\cdot \vec{v}_2\big)
 \big(\vec{S}_1\cdot \vec{S}_2\big)  \big(\vec{v}_1\cdot \vec{n}\big)+2 \big(\vec{S}_1\cdot \vec{v}_1\big) \big(\vec{S}_1\cdot \vec{v}_2\big)\big(\vec{S}_1\cdot \vec{S}_2\big)
\nonumber\\
&\quad   -15 S_1^2 \big(\vec{S}_1\cdot \vec{n}\big)\big(\vec{S}_2\cdot
\vec{v}_1\big)\big(\vec{v}_2\cdot \vec{n}\big) -15 S_1^2 \big(\vec{S}_1\cdot \vec{n}\big)\big(\vec{S}_2\cdot \vec{v}_2\big)\big(\vec{v}_1\cdot \vec{n}\big) 
 \nonumber\\
&\quad -15 S_1^2 \big(\vec{S}_1\cdot \vec{n}\big) \big(\vec{S}_2\cdot \vec{n}\big)  \big(\vec{v}_1\cdot \vec{v}_2\big) +105 S_1^2 \big(\vec{S}_1\cdot \vec{n}\big)\big(\vec{S}_2\cdot \vec{n}\big) \big(\vec{v}_1\cdot \vec{n}\big)
\big(\vec{v}_2\cdot \vec{n}\big) \nonumber\\
&\quad -15 S_1^2 \big(\vec{S}_1\cdot \vec{v}_1\big) \big(\vec{S}_2\cdot\vec{n}\big) \big(\vec{v}_2\cdot \vec{n}\big)   -15 S_1^2\big(\vec{S}_1\cdot\vec{v}_2\big)
\big(\vec{S}_2\cdot \vec{n}\big) \big(\vec{v}_1\cdot \vec{n}\big)  \nonumber\\
&\quad +3
S_1^2 \big(\vec{S}_1\cdot \vec{v}_2\big) \big(\vec{S}_2\cdot \vec{v}_1\big)-5 S_1^2 \big(\vec{S}_1\cdot \vec{S}_2\big) \big(\vec{v}_1\cdot \vec{n}\big)
\big(\vec{v}_2\cdot \vec{n}\big) \nonumber\\
&\quad +3 S_1^2 \big(\vec{S}_1\cdot \vec{v}_1\big) \big(\vec{S}_2\cdot \vec{v}_2\big)+S_1^2
\big(\vec{S}_1\cdot \vec{S}_2\big) \big(\vec{v}_1\cdot \vec{v}_2\big)
\Big]\nonumber\\
&\quad +\frac{C_{1(\text{BS}^3)}}{4}\frac{G}{m_1^2r^4}\Big[5 \big(\vec{S}_1\cdot \vec{n}\big)^3
\big(\dot{\vec{S}}_2\cdot \vec{v}_1\big) 
-35 \big(\vec{S}_1\cdot \vec{n}\big)^3 \big(\dot{\vec{S}}_2\cdot\vec{n}\big) \big(\vec{v}_1\cdot \vec{n}\big) \nonumber\\
&\quad +105\big(\vec{S}_1\cdot \vec{n}\big)^2 \big(\dot{\vec{S}}_1\cdot \vec{n}\big)
\big(\vec{S}_2\cdot \vec{n}\big) \big(\vec{v}_2\cdot \vec{n}\big) -5 \big(\vec{S}_1\cdot \vec{n}\big)^2 \big(\dot{\vec{S}}_1\cdot \vec{S}_2\big) \big(\vec{v}_2\cdot\vec{n}\big)\nonumber\\
&\quad -15 \big(\vec{S}_1\cdot \vec{n}\big)^2
\big(\dot{\vec{S}}_1\cdot \vec{v}_2\big) \big(\vec{S}_2\cdot \vec{n}\big)+5 \big(\vec{S}_1\cdot \vec{n}\big)^2
\big(\vec{S}_1\cdot\dot{\vec{S}}_2\big) \big(\vec{v}_1\cdot \vec{n}\big)\nonumber\\
&\quad +15 \big(\vec{S}_1\cdot \vec{n}\big)^2\big(\vec{S}_1\cdot \vec{v}_1\big)\big(\dot{\vec{S}}_2\cdot \vec{n}\big)
 -15 \big(\vec{S}_1\cdot \vec{n}\big)^2 \big(\dot{\vec{S}}_1\cdot \vec{n}\big)
\big(\vec{S}_2\cdot \vec{v}_2\big)\nonumber\\
&\quad -30 \big(\vec{S}_1\cdot \vec{n}\big)\big(\dot{\vec{S}}_1\cdot \vec{S}_1\big)\big(\vec{S}_2\cdot \vec{n}\big) \big(\vec{v}_2\cdot\vec{n}\big)   -10\big(\vec{S}_1\cdot\vec{n}\big)
\big(\dot{\vec{S}}_1\cdot \vec{n}\big) \big(\vec{S}_1\cdot \vec{S}_2\big)\big(\vec{v}_2\cdot \vec{n}\big) \nonumber\\
&\quad +2 \big(\vec{S}_1\cdot \vec{n}\big)\big(\dot{\vec{S}}_1\cdot \vec{v}_2\big) \big(\vec{S}_1\cdot \vec{S}_2\big) -2 \big(\vec{S}_1\cdot \vec{n}\big)\big(\vec{S}_1\cdot \vec{v}_1\big)
\big(\vec{S}_1\cdot\dot{\vec{S}}_2\big) \nonumber\\
&\quad -30\big(\vec{S}_1\cdot\vec{n}\big)\big(\vec{S}_1\cdot \vec{v}_2\big)
\big(\dot{\vec{S}}_1\cdot \vec{n}\big) \big(\vec{S}_2\cdot \vec{n}\big)  +2\big(\vec{S}_1\cdot \vec{n}\big) \big(\vec{S}_1\cdot \vec{v}_2\big)\big(\dot{\vec{S}}_1\cdot \vec{S}_2\big)  \nonumber\\
&\quad +6\big(\vec{S}_1\cdot \vec{n}\big)
\big(\dot{\vec{S}}_1\cdot \vec{S}_1\big) \big(\vec{S}_2\cdot \vec{v}_2\big) -15 S_1^2
\big(\dot{\vec{S}}_1\cdot \vec{n}\big) \big(\vec{S}_2\cdot \vec{n}\big) \big(\vec{v}_2\cdot \vec{n}\big) \nonumber\\
&\quad -3 S_1^2 \big(\vec{S}_1\cdot \vec{n}\big)
\big(\dot{\vec{S}}_2\cdot \vec{v}_1\big) +15 S_1^2 \big(\vec{S}_1\cdot \vec{n}\big) \big(\dot{\vec{S}}_2\cdot\vec{n}\big) \big(\vec{v}_1\cdot \vec{n}\big)  \nonumber\\
&\quad + S_1^2  \big(\dot{\vec{S}}_1\cdot \vec{S}_2\big) \big(\vec{v}_2\cdot \vec{n}\big) +3 S_1^2 \big(\dot{\vec{S}}_1\cdot \vec{v}_2\big)
\big(\vec{S}_2\cdot \vec{n}\big) - S_1^2 \big(\vec{S}_1\cdot\dot{\vec{S}}_2\big)\big(\vec{v}_1\cdot\vec{n}\big)\nonumber\\
&\quad -3 S_1^2 \big(\vec{S}_1\cdot \vec{v}_1\big) \big(\dot{\vec{S}}_2\cdot \vec{n}\big)+3 S_1^2 \big(\dot{\vec{S}}_1\cdot \vec{n}\big)
\big(\vec{S}_2\cdot \vec{v}_2\big)+2 
\big(\dot{\vec{S}}_1\cdot \vec{S}_1\big) \big(\vec{S}_1\cdot \vec{S}_2\big)\big(\vec{v}_2\cdot \vec{n}\big)\nonumber\\
&\quad 
 +6 \big(\vec{S}_1\cdot \vec{v}_2\big)\big(\dot{\vec{S}}_1\cdot \vec{S}_1\big)\big(\vec{S}_2\cdot \vec{n}\big) +2 \big(\vec{S}_1\cdot \vec{v}_2\big)\big(\dot{\vec{S}}_1\cdot \vec{n}\big)
\big(\vec{S}_1\cdot \vec{S}_2\big) 
\Big]\nonumber\\
&\quad+\frac{C_{1(\text{BS}^3)}}{12}\frac{G}{m_1^2r^3}\Big[45 \big(\vec{S}_1\cdot\vec{n}\big)^2\big(\dot{\vec{S}}_1\cdot \vec{n}\big) \big(\dot{\vec{S}}_2\cdot \vec{n}\big) 
-3 \big(\vec{S}_1\cdot \vec{n}\big)^2\big(\dot{\vec{S}}_1\cdot \dot{\vec{S}}_2\big) \nonumber\\
&\quad -18 \big(\vec{S}_1\cdot\vec{n}\big) \big(\dot{\vec{S}}_1\cdot \vec{S}_1\big) \big(\dot{\vec{S}}_2\cdot \vec{n}\big) -6\big(\vec{S}_1\cdot\vec{n}\big)\big(\dot{\vec{S}}_1\cdot \vec{n}\big) \big(\vec{S}_1\cdot\dot{\vec{S}}_2\big)  \nonumber\\
&\quad +2 \big(\dot{\vec{S}}_1\cdot \vec{S}_1\big) \big(\vec{S}_1\cdot\dot{\vec{S}}_2\big)-9 S_1^2\big(\dot{\vec{S}}_1\cdot \vec{n}\big) \big(\dot{\vec{S}}_2\cdot \vec{n}\big)+S_1^2\big(\dot{\vec{S}}_1\cdot
\dot{\vec{S}}_2\big) 
\Big] ,&\\
\text{Fig.~2(b3)}&=2C_{1(\text{BS}^3)}\frac{G}{m_1^2 r^5} \Bigg\{35 \big( \vec{S}_1\cdot\vec{n}\big)^3\big(\vec{S}_2\cdot \vec{v}_1\big) \big( \vec{v}_2\cdot\vec{n}\big) -35 \big(\vec{S}_1\cdot\vec{n}\big)^3\big(\vec{S}_2\cdot\vec{n} \big)\big(\vec{v}_1\cdot \vec{v}_2\big)\nonumber\\*
& -35 \big(\vec{S}_1\cdot \vec{S}_2\big) \big(\vec{S}_1\cdot\vec{n}\big)^2 \big( \vec{v}_1\cdot\vec{n}\big)\big(\vec{v}_2\cdot\vec{n}\big)+35 \big(\vec{S}_1\cdot\vec{n}\big)^2\big(\vec{S}_1\cdot \vec{v}_2\big)\big(\vec{S}_2\cdot\vec{n}\big) \big(\vec{v}_1\cdot\vec{n}\big)\nonumber\\*
& -20 \big(\vec{S}_1\cdot\vec{n}\big)^2 \big(\vec{S}_1\cdot \vec{v}_2\big)\big( \vec{S}_2\cdot \vec{v}_1\big)+20 \big(\vec{S}_1\cdot\vec{n}\big)^2\big(\vec{S}_1\cdot \vec{S}_2\big)\big(\vec{v}_1\cdot \vec{v}_2\big)\nonumber\\*
& +10 \big(\vec{S}_1\cdot\vec{n}\big)\big(\vec{S}_1\cdot\vec{v}_1\big)\big(\vec{S}_1\cdot \vec{S}_2\big)  \big(\vec{v}_2\cdot\vec{n}\big)-10  \big(\vec{S}_1\cdot\vec{n}\big)\big(\vec{S}_1\cdot\vec{v}_1\big)\big(\vec{S}_1\cdot \vec{v}_2\big)\big(\vec{S}_2\cdot\vec{n}\big)\nonumber\\
& +S_1^2\Big[15   \big( \vec{S}_1\cdot\vec{n}\big)\big(\vec{S}_2\cdot\vec{n}\big)\big(\vec{v}_1\cdot \vec{v}_2\big)-15\big(\vec{S}_1\cdot\vec{n}\big)\big(\vec{S}_2\cdot \vec{v}_1\big)\big(\vec{v}_2\cdot\vec{n}\big)-4 \big(\vec{S}_1\cdot\vec{S}_2\big)\big(\vec{v}_1\cdot\vec{v}_2\big)\nonumber\\
& +5 \big(\vec{S}_1\cdot \vec{S}_2\big) \big(\vec{v}_1\cdot \vec{n}\big)\big(\vec{v}_2\cdot \vec{n}\big) -5  \big(\vec{S}_1\cdot\vec{v}_2\big)\big(\vec{S}_2\cdot\vec{n}\big)\big(\vec{v}_1\cdot\vec{n}\big)+4 \big(\vec{S}_1\cdot\vec{v}_2\big)\big(\vec{S}_2\cdot\vec{v}_1\big)\Big]\Bigg\},
&\\
\text{Fig.~2(b4)}&=\frac{C_{1(\text{BS}^3)}}{2}\frac{G}{m_1^2r^5}\Big[
35 \big(
\vec{S}_1\cdot\vec{n}\big)^3  \big(\vec{S}_2\cdot \vec{v}_2\big)\big(\vec{v}_1\cdot \vec{n}\big)-70
\big(\vec{S}_1\cdot \vec{n}\big)^3  \big(\vec{S}_2\cdot \vec{v}_1\big)\big(\vec{v}_2\cdot\vec{n}\big)\nonumber\\
&\quad -70 \big(\vec{S}_1\cdot\vec{n}\big){}^2 \big(\vec{S}_1\cdot
\vec{v}_2\big) \big(\vec{S}_2\cdot\vec{n}\big) \big(\vec{v}_1\cdot\vec{n}\big) +35 \big(\vec{S}_1\cdot\vec{n}\big){}^2 \big(\vec{S}_1\cdot \vec{v}_1\big)\big(\vec{S}_2\cdot\vec{n}\big) \big(\vec{v}_2\cdot\vec{n}\big)
\nonumber\\
&\quad +10 \big(\vec{S}_1\cdot\vec{n}\big)  \big(\vec{S}_1\cdot
\vec{v}_1\big) \big(\vec{S}_1\cdot \vec{v}_2\big)\big(\vec{S}_2\cdot\vec{n}\big)+35 \big(\vec{S}_1\cdot\vec{n}\big){}^2 \big(\vec{S}_1\cdot \vec{S}_2\big) \big(\vec{v}_1\cdot\vec{n}\big)
\big(\vec{v}_2\cdot\vec{n}\big) \nonumber\\
&\quad -15 S_1^2 \big(\vec{S}_1\cdot\vec{n}\big)  \big(\vec{S}_2\cdot \vec{v}_2\big)\big(\vec{v}_1\cdot\vec{n}\big)+10 \big(\vec{S}_1\cdot \vec{n}\big)\big(\vec{S}_1\cdot \vec{v}_2\big)
\big(\vec{S}_1\cdot \vec{S}_2\big) \big(\vec{v}_1\cdot \vec{n}\big)\nonumber\\
&\quad +30 S_1^2 \big(\vec{S}_1\cdot \vec{n}\big)
 \big(\vec{S}_2\cdot \vec{v}_1\big)\big(\vec{v}_2\cdot \vec{n}\big) -20 \big(\vec{S}_1\cdot\vec{n}\big) \big(\vec{S}_1\cdot \vec{v}_1\big) \big(\vec{S}_1\cdot \vec{S}_2\big) \big(\vec{v}_2\cdot \vec{n}\big)\nonumber\\
&\quad -5 \big(\vec{S}_1\cdot\vec{n}\big){}^2 \big(\vec{S}_1\cdot \vec{S}_2\big) \big(\vec{v}_1\cdot \vec{v}_2\big)-20 \big(\vec{S}_1\cdot \vec{n}\big){}^2
\big(\vec{S}_1\cdot \vec{v}_1\big) \big(\vec{S}_2\cdot \vec{v}_2\big)\nonumber\\
&\quad +40 \big(\vec{S}_1\cdot \vec{n}\big){}^2
\big(\vec{S}_1\cdot \vec{v}_2\big) \big(\vec{S}_2\cdot \vec{v}_1\big)+10S_1^2\big(\vec{S}_1\cdot \vec{v}_2\big) \big(\vec{S}_2\cdot \vec{n}\big)  \big(\vec{v}_1\cdot\vec{n}\big) \nonumber\\
&\quad -5 S_1^2\big(\vec{S}_1\cdot \vec{v}_1\big)\big(\vec{S}_2\cdot \vec{n}\big)\big(\vec{v}_2\cdot \vec{n}\big)-5 S_1^2  \big(\vec{S}_1\cdot
\vec{S}_2\big)\big(\vec{v}_1\cdot\vec{n}\big) \big(\vec{v}_2\cdot \vec{n}\big)\nonumber\\
&\quad +S_1^2 \big(\vec{S}_1\cdot \vec{S}_2\big) \big(\vec{v}_1\cdot \vec{v}_2\big)+4
S_1^2 \big(\vec{S}_1\cdot \vec{v}_1\big) \big(\vec{S}_2\cdot \vec{v}_2\big)\nonumber\\
&\quad -8 S_1^2 \big(\vec{S}_1\cdot \vec{v}_2\big) \big(\vec{S}_2\cdot \vec{v}_1\big)
\Big]+\frac{C_{1(\text{BS}^3)}}{2}\frac{G}{m_1^2r^4}\Big[
5
\big(\dot{\vec{S}}_1\cdot \vec{S}_2\big) \big(\vec{S}_1\cdot \vec{n}\big){}^2 \big(\vec{v}_2\cdot \vec{n}\big)\nonumber\\
&\quad +8 \big(\dot{\vec{S}}_1\cdot \vec{S}_1\big) \big(\vec{S}_1\cdot \vec{n}\big) \big(\vec{S}_2\cdot \vec{v}_2\big)-2
\big(\dot{\vec{S}}_1\cdot \vec{S}_1\big) \big(\vec{S}_1\cdot \vec{v}_2\big)\big(\vec{S}_2\cdot \vec{n}\big) \nonumber\\
&\quad -10 \big(\dot{\vec{S}}_1\cdot \vec{S}_1\big) \big(\vec{S}_1\cdot \vec{n}\big) \big(\vec{S}_2\cdot \vec{n}\big) \big(\vec{v}_2\cdot\vec{n}\big)-6
\big(\dot{\vec{S}}_1\cdot \vec{S}_2\big) \big(\vec{S}_1\cdot \vec{n}\big) \big(\vec{S}_1\cdot \vec{v}_2\big)\nonumber\\
&\quad +10
\big(\dot{\vec{S}}_1\cdot \vec{v}_2\big) \big(\vec{S}_1\cdot \vec{n}\big){}^2 \big(\vec{S}_2\cdot \vec{n}\big)-6
\big(\dot{\vec{S}}_1\cdot \vec{v}_2\big) \big(\vec{S}_1\cdot \vec{n}\big) \big(\vec{S}_1\cdot
\vec{S}_2\big)\nonumber\\
&\quad  +20
\big(\dot{\vec{S}}_1\cdot\vec{n}\big)\big(\vec{S}_1\cdot
\vec{v}_2\big) \big(\vec{S}_1\cdot \vec{n}\big) \big(\vec{S}_2\cdot \vec{n}\big) -6 \big(\dot{\vec{S}}_1\cdot\vec{n}\big)\big(\vec{S}_1\cdot \vec{v}_2\big) \big(\vec{S}_1\cdot \vec{S}_2\big)\nonumber\\
&\quad +10 \big(\dot{\vec{S}}_1\cdot\vec{n}\big) \big(\vec{S}_1\cdot \vec{n}\big) 
\big(\vec{S}_1\cdot \vec{S}_2\big)\big(\vec{v}_2\cdot \vec{n}\big)-15 \big(\dot{\vec{S}}_1\cdot\vec{n}\big) \big(\vec{S}_1\cdot \vec{n}\big){}^2
\big(\vec{S}_2\cdot \vec{v}_2\big)\nonumber\\
&\quad -5 S_1^2\big(\dot{\vec{S}}_1\cdot\vec{n}\big) \big(\vec{S}_2\cdot \vec{n}\big) \big(\vec{v}_2\cdot\vec{n}\big) -S_1^2\big(\dot{\vec{S}}_1\cdot \vec{v}_2\big) \big(\vec{S}_2\cdot \vec{n}\big)+4 S_1^2\big(\dot{\vec{S}}_1\cdot\vec{n}\big) 
\big(\vec{S}_2\cdot \vec{v}_2\big)
\Big] ,&\\
\text{Fig.~2(c1)}&= \frac{C_{1(\text{ES}^4)}}{16}\frac{Gm_2}{m_1^3r^5}\Bigg\{\Big[35\big(\vec{S}_1\cdot\vec{n}\big)^4-30S_1^2\big(\vec{S}_1\cdot\vec{n}\big)^2+3S_1^4\Big](2+3v_1^2+3v_2^2)\nonumber\\*
&\quad -140 \big(\vec{S}_1\cdot\vec{n}\big)^3\big(\vec{S}_1\cdot \vec{v}_1\big) \big(\vec{v}_1\cdot\vec{n}\big) +60 \big(\vec{S}_1\cdot\vec{n}\big)^2 \big(\vec{S}_1\cdot \vec{v}_1\big)^2-12 S_1^2 \big(\vec{S}_1\cdot \vec{v}_1\big)^2\nonumber\\*
&\quad +60 S_1^2
\big(\vec{S}_1\cdot\vec{n}\big)\big( \vec{S}_1\cdot \vec{v}_1\big) \big(\vec{v}_1\cdot\vec{n}\big)\Bigg\}+\frac{C_{1(\text{ES}^4)}}{8}\frac{Gm_2}{m_1^3r^4}\Big[5S_1^2\big(\vec{S}_1\cdot\vec{n}\big)^2\big(\vec{a}_1\cdot\vec{n}\big)\nonumber\\*
&\quad -2S_1^2\big(\vec{S}_1\cdot\vec{n}\big)\big(\vec{S}_1\cdot\vec{a}_1\big)-S_1^4\big(\vec{a}_1\cdot\vec{n}\big)\Big]+\frac{C_{1(\text{ES}^4)}}{12}\frac{Gm_2}{m_1^3r^3}\Big[3 \big(\ddot{\vec{S}}_1\cdot \vec{S}_1\big) \big( \vec{S}_1\cdot\vec{n}\big)^2\nonumber\\*
&\quad +3 S_1^2 \big( \ddot{\vec{S}}_1\cdot\vec{n}\big) \big( \vec{S}_1\cdot\vec{n}\big)-2 S_1^2 \big(\ddot{\vec{S}}_1\cdot \vec{S}_1\big)+3
S_1^2 \big( \dot{\vec{S}}_1\cdot\vec{n}\big)^2+3 \dot{S}_1^2 \big(\vec{S}_1\cdot\vec{n}\big)^2\nonumber\\*
&\quad +12 \big(\dot{\vec{S}}_1\cdot\vec{n}\big) \big(\dot{\vec{S}}_1\cdot \vec{S}_1\big) \big(\vec{S}_1\cdot\vec{n}\big)-2 \dot{S}_1^2 S_1^2-4 \big(\dot{\vec{S}}_1\cdot \vec{S}_1\big)^2\Big],&\\
\text{Fig.~2(c2)}&= -\frac{C_{1(\text{ES}^4)}}{16}\frac{Gm_2}{m_1^3r^5}\Big[315(\vec{S}_1\cdot\vec{n})^4(\vec{v}_1\cdot\vec{n})(\vec{v}_2\cdot\vec{n})-210S_1^2(\vec{S}_1\cdot \vec{n})^2(\vec{v}_1\cdot\vec{n})(\vec{v}_2\cdot\vec{n})\nonumber\\
&+15S_1^4(\vec{v}_1\cdot\vec{n})(\vec{v}_2\cdot\vec{n})-140(\vec{S}_1\cdot\vec{n})^3(\vec{S}_1\cdot\vec{v}_1)(\vec{v}_2\cdot\vec{n})+60S_1^2(\vec{S}_1\cdot\vec{v}_1)(\vec{S}_1\cdot\vec{n})(\vec{v}_2\cdot\vec{n})\nonumber\\
&-140(\vec{S}_1\cdot\vec{n})^3(\vec{S}_1\cdot\vec{v}_2)(\vec{v}_1\cdot\vec{n})+60S_1^2(\vec{S}_1\cdot\vec{n})(\vec{S}_1\cdot\vec{v}_2)(\vec{v}_1\cdot\vec{n})-35(\vec{S}_1\cdot\vec{n})^4(\vec{v}_1\cdot\vec{v}_2)\nonumber\\
&+60(\vec{S}_1\cdot\vec{n})^2(\vec{S}_1\cdot\vec{v}_1)(\vec{S}_1\cdot\vec{v}_2)-12S_1^2(\vec{S}_1\cdot\vec{v}_1)(\vec{S}_1\cdot\vec{v}_2)+30S_1^2(\vec{S}_1\cdot\vec{n})^2(\vec{v}_1\cdot\vec{v}_2)\nonumber\\
&-3S_1^4(\vec{v}_1\cdot\vec{v}_2)\Big]-\frac{C_{1(\text{ES}^4)}}{4}\frac{Gm_2}{m_1^3r^4}\Big[5(\dot{\vec{S}}_1\cdot\vec{v}_2)(\vec{S}_1\cdot\vec{n})^3+15(\dot{\vec{S}}_1\cdot\vec{S}_1)(\vec{S}_1\cdot\vec{n})^2(\vec{v}_2\cdot\vec{n})\nonumber\\
&-35(\dot{\vec{S}}_1\cdot\vec{n})(\vec{S}_1\cdot\vec{n})^3(\vec{v}_2\cdot\vec{n})-3S_1^2(\dot{\vec{S}}_1\cdot\vec{v}_2)(\vec{S}_1\cdot\vec{n})-3S_1^2(\dot{\vec{S}}_1\cdot\vec{S}_1)(\vec{v}_2\cdot\vec{n})\nonumber\\
&+15S_1^2(\dot{\vec{S}}_1\cdot\vec{n})(\vec{S}_1\cdot\vec{n})(\vec{v}_2\cdot\vec{n})-6(\dot{\vec{S}}_1\cdot\vec{S}_1)(\vec{S}_1\cdot\vec{v}_2)(\vec{S}_1\cdot\vec{n})\nonumber\\
&+15(\dot{\vec{S}}_1\cdot\vec{n})(\vec{S}_1\cdot\vec{n})^2(\vec{S}_1\cdot\vec{v}_2)-3S_1^2(\dot{\vec{S}}_1\cdot\vec{n})(\vec{S}_1\cdot\vec{v}_2)\Big],&\\
\text{Fig.~2(c3)}&=
\frac{ C_{1(\text{ES}^4)}}{2}\frac{G m_2}{m_1^3 r^5}
	\big(\vec{v}_1\cdot \vec{v}_2\big)\Big[30 S_1^2  \big(\vec{S}_1\cdot\vec{n}\big)^2-35\big( \vec{S}_1\cdot\vec{n}\big)^4-3S_1^4\Big]
\nonumber\\*
&\quad
+\frac{C_{1(\text{ES}^4)}}{2}\frac{G m_2 }{m_1^3 r^4} \Big[5 \big(\dot{\vec{S}}_1\cdot \vec{v}_2\big) \big(\vec{S}_1\cdot \vec{n}\big)^3+15 \big( \dot{\vec{S}}_1\cdot\vec{n}\big)
	\big(n\cdot S_1\big)^2 \big(\vec{S}_1\cdot \vec{v}_2\big)\nonumber\\*
	&\quad-3S_1^2 \big(\dot{\vec{S}}_1\cdot \vec{v}_2\big) \big(\vec{S}_1\cdot\vec{n}\big)-6 \big(\dot{\vec{S}}_1\cdot \vec{S}_1\big)  \big(\vec{S}_1\cdot\vec{n}\big)
	\big(\vec{S}_1\cdot \vec{v}_2\big)-3S_1^2 \big(\dot{\vec{S}}_1\cdot\vec{n}\big) \big(\vec{S}_1\cdot \vec{v}_2\big)\Big].&
\end{align}
Notice that almost all these graphs contain higher-order time derivative terms, notably
even at second order, whereas at the LO no higher-order time derivatives 
appeared yet~\cite{Levi:2014gsa}.

\subsection{Two-graviton exchange}

\begin{figure}[t]
\centering
\includegraphics[scale=0.5]{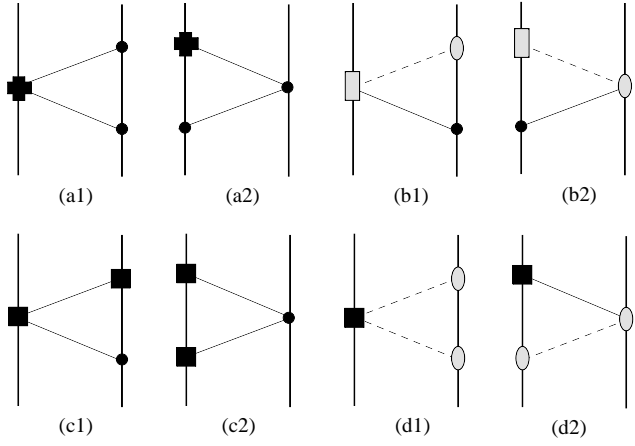}
\caption{The Feynman graphs of two-graviton exchange at the NLO quartic-in-spin interaction. 
These graphs contain all relevant interactions among the mass, spin and spin-induced 
multipoles up to the hexadecapole. In particular at this nonlinear level there are also 
interactions with the various multipoles on two different points of the worldline, which 
occurs as of the NLO spin-squared sector \cite{Levi:2015msa,Levi:2015ixa,Levi:2019kgk}, such 
as a spin dipole and a spin-induced quadrupole or two spin-induced quadrupoles on the same 
worldline. The graph (a1) contains a new two-graviton coupling to the hexadecapole.}
\label{s4nlo2g}
\end{figure}

There are $8$ graphs of two-graviton exchange in this sector, shown in figure \ref{s4nlo2g}, 
none of which contains time derivatives. The graph 3(a1) contains a new two-graviton coupling 
to the hexadecapole.

The graphs in figure \ref{s4nlo2g} have the following values:
\begin{align}
\text{Fig.~3(a1)}&=-\frac{3C_{1(\text{ES}^4)}}{8}\frac{G^2m_2^2}{m_1^3r^6}
\left[95\big(\vec{S}_1\cdot\vec{n}\big)^4-81S_1^2\big(\vec{S}_1\cdot\vec{n}\big)^2+8S_1^4\right] ,&\\
\text{Fig.~3(a2)}&=-\frac{C_{1(\text{ES}^4)}}{8}\frac{G^2m_2}{m_1^2r^6}
\left[3S_1^4-30S_1^2\big(\vec{S}_1\cdot\vec{n}\big)^2+35\big(\vec{S}_1\cdot\vec{n}\big)^4\right] ,&\\
\text{Fig.~3(b1)}&=-\frac{C_{1(\text{BS}^3)}}{3}\frac{G^2m_2}{m_1^2r^6}
\left[441\big(\vec{S}_1\cdot\vec{n}\big)^3\big(\vec{S}_2\cdot\vec{n}\big)
-189S_1^2\big(\vec{S}_1\cdot\vec{n}\big)\big(\vec{S}_2\cdot\vec{n}\big)\right.\nonumber\\
&\qquad\qquad\qquad\qquad\;\left.
-183\big(\vec{S}_1\cdot\vec{S}_2\big)\big(\vec{S}_1\cdot\vec{n}\big)^2
+35S_1^2\big(\vec{S}_1\cdot\vec{S}_2\big)\right],&\\
\text{Fig.~3(b2)}&=-2C_{1(\text{BS}^3)}\frac{G^2}{m_1r^6}
\left[3S_1^2\big(\vec{S}_1\cdot\vec{S}_2\big)
-15S_1^2\big(\vec{S}_1\cdot\vec{n}\big)\big(\vec{S}_2\cdot\vec{n}\big)
-15\big(\vec{S}_1\cdot\vec{S}_2\big)\big(\vec{S}_1\cdot\vec{n}\big)^2\right.\nonumber\\*
&\qquad\qquad\qquad\qquad\left.
+35\big(\vec{S}_1\cdot\vec{n}\big)^3\big(\vec{S}_2\cdot\vec{n}\big)\right] ,&\\
\text{Fig.~3(c1)}&=-\frac{9C_{1(\text{ES}^2)}C_{2(\text{ES}^2)}}{2}\frac{G^2}{m_1r^6}
\left[S_1^2S_2^2-4S_1^2\big(\vec{S}_2\cdot\vec{n}\big)
-4S_2^2\big(\vec{S}_1\cdot\vec{n}\big)^2+\big(\vec{S}_1\cdot\vec{S}_2\big)^2\right.\nonumber\\
&\qquad\qquad\qquad\quad\left.
-12\big(\vec{S}_1\cdot\vec{S}_2\big)\big(\vec{S}_1\cdot\vec{n}\big)\big(\vec{S}_2\cdot\vec{n}\big)
+24\big(\vec{S}_1\cdot\vec{n}\big)^2\big(\vec{S}_2\cdot\vec{n}\big)^2\right] ,&\\
\text{Fig.~3(c2)}&=-\frac{C_{1(\text{ES}^2)}^2}{8}\frac{G^2m_2}{m_1^2r^6}
\left[3\big(\vec{S}_1\cdot\vec{n}\big)^2-S_1^2\right]^{2} ,&\\
\text{Fig.~3(d1)}&=-\frac{C_{1(\text{ES}^2)}}{2}\frac{G^2}{m_1r^6}
\left[S_1^2S_2^2+\big(\vec{S}_1\cdot\vec{S}_2\big)^2+3S_1^2\big(\vec{S}_2\cdot\vec{n}\big)^2
+9\big(\vec{S}_1\cdot\vec{n}\big)^2\big(\vec{S}_2\cdot\vec{n}\big)^2\right.\nonumber\\*
&\qquad\qquad\qquad\qquad\left.
-6\big(\vec{S}_1\cdot\vec{n}\big)\big(\vec{S}_1\cdot\vec{S}_2\big)\big(\vec{S}_2\cdot\vec{n}\big)\right] ,&\\
\text{Fig.~3(d2)}&=-2C_{1(\text{ES}^2)}\frac{G^2}{m_1r^6}
\left[3\big(\vec{S}_1\cdot\vec{n}\big)^2-S_1^2\right]
\left[3\big(\vec{S}_1\cdot\vec{n}\big)\big(\vec{S}_2\cdot\vec{n}\big)-\big(\vec{S}_1\cdot\vec{S}_2\big)\right] .&
\end{align}

\subsection{Cubic self-interaction}

\begin{figure}[t]
\centering
\includegraphics[scale=0.7]{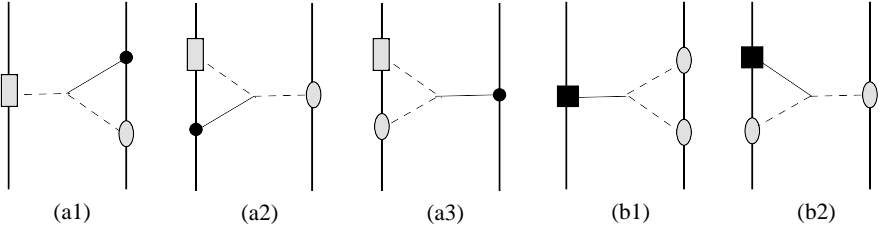}
\caption{The Feynman graphs with cubic self-gravitational interaction, i.e.~at one-loop 
level, at the NLO quartic-in-spin interaction. There are no vertices with time dependence 
here, similar to the NLO quadratic-in-spin sector of even-parity \cite{Levi:2015msa}.
These graphs contain all possible interactions among the mass, spin and spin-induced 
multipoles up to the hexadecapole.}
\label{s4nlo1loop}
\end{figure}

There are $5$ graphs of cubic self-interaction in this sector, shown in figure 
\ref{s4nlo1loop}, none of which contains time-dependent
self-interaction, similar to the NLO quadratic-in-spin sector of even parity
\cite{Levi:2015msa}. These graphs contain all possible interactions among the mass, spin and 
spin-induced multipoles up to the hexadecapole, similar to the nonlinear graphs of 
two-graviton exchange.

The graphs in figure \ref{s4nlo1loop} have the following values:
\begin{align}
\text{Fig.~4(a1)}&=-\frac{4C_{1(\text{BS}^3)}}{3}\frac{G^2m_2}{m_1^2r^6}
\Big[9S_1^2(\vec{S}_1\cdot\vec{n})(\vec{S}_2\cdot\vec{n})-24(\vec{S}_1\cdot\vec{n})^3(\vec{S}_2\cdot\vec{n})\nonumber\\
&\qquad\qquad\qquad\qquad\quad
+12(\vec{S}_1\cdot\vec{n})^2(\vec{S}_1\cdot\vec{S}_2)-2S_1^2(\vec{S}_1\cdot\vec{S}_2)\Big] ,&\\
\text{Fig.~4(a2)}&=-3C_{1(\text{BS}^3)}\frac{G^2}{m_1r^6}
\Big[4S_1^2(\vec{S}_1\cdot\vec{n})(\vec{S}_2\cdot\vec{n})-10(\vec{S}_1\cdot\vec{n})^3(\vec{S}_2\cdot\vec{n})\nonumber\\*
&\qquad\qquad\qquad\qquad+5(\vec{S}_1\cdot\vec{n})^2(\vec{S}_1\cdot\vec{S}_2)-S_1^2(\vec{S}_1\cdot\vec{S}_2)\Big] ,&\\
\text{Fig.~4(a3)}&=C_{1(\text{BS}^3)}\frac{G^2m_2}{m_1^2r^6}(\vec{S}_1\cdot\vec{n})^2
\Big[3S_1^2-5(\vec{S}_1\cdot\vec{n})^2\Big] ,&\\
\text{Fig.~4(b1)}&=-C_{1(\text{ES}^2)}\frac{G^2}{m_1r^6}
\Big[24(\vec{S}_1\cdot\vec{n})^2(\vec{S}_2\cdot\vec{n})^2
-6S_1^2(\vec{S}_2\cdot\vec{n})^2+(\vec{S}_1\cdot\vec{S}_2)^2-S_1^2S_2^2\nonumber\\*
&\qquad\qquad\qquad\qquad
-12(\vec{S}_1\cdot\vec{n})(\vec{S}_1\cdot\vec{S}_2)(\vec{S}_2\cdot\vec{n})\Big] ,&\\
\text{Fig.~4(b2)}&=-C_{1(\text{ES}^2)}\frac{G^2}{m_1r^6}
\Big[6(\vec{S}_1\cdot\vec{n})^3(\vec{S}_2\cdot\vec{n})
-3(\vec{S}_1\cdot\vec{n})^2(\vec{S}_1\cdot\vec{S}_2)-S_1^2(\vec{S}_1\cdot\vec{S}_2)\Big] .&
\end{align}

\section{Composite worldline couplings}
\label{newfromgauge?}

The formulation of the EFT of a spinning particle in \cite{Levi:2015msa} assumed 
an initial covariant gauge of the rotational DOFs in terms of the linear momentum, as 
originally put forward by Tulczyjew \cite{Tulczyjew:1959b} (extended by Dixon 
\cite{Dixon:1970zza} later to higher-multipoles), which was proven to be uniquely 
distinguished \cite{Schattner:1979vn,Schattner:1979vp}. As detailed in section 4 of 
\cite{Levi:2019kgk}, and pointed out already in \cite{Levi:2015msa}, this gives rise to 
composite worldline couplings in sectors of higher-spin as of the NLO cubic-in-spin sector as 
the linear momentum can no longer be considered independent of the spin
\be
p_{\mu} = -\frac{\partial L}{\partial u^{\mu}}
= m \frac{u_{\mu}}{\sqrt{u}^2} + {\cal{O}}(RS^2).
\ee
The correction to the linear momentum which was already required for the NLO cubic-in-spin 
sector \cite{Levi:2019kgk} is given by
\begin{align} \label{delp1}
\Delta p_{\kappa}[S^2] \equiv p_{\kappa}[S^2]-\bar{p}_{\kappa}
\simeq \frac{C_{ES^2}}{2m} S^\mu S^\nu 
\left(\frac{2}{u} R_{\mu \alpha \nu \kappa} u^{\alpha}
-\frac{1}{u^3} R_{\mu \alpha \nu \beta} u^{\alpha} u^{\beta} u_{\kappa}
\right),
\end{align}
where $\bar{p}_{\kappa}\equiv \tfrac{m}{u} u_{\kappa}$ is the leading approximation to the 
linear momentum. 
At this order it is clear from using eq.~(4.8) of \cite{Levi:2015msa} that the spin-induced 
multipole itself, rewritten as a contraction of spin tensors, has no $u^{\mu}$ dependence and 
thus no additional contribution to the linear momentum.

At the current NLO quartic-in-spin sector we also have to consider the next spin-dependent 
correction to the linear momentum. Similar to the correction in eq.~\eqref{delp1}, by first 
treating the product of spin vectors as independent of $u^{\mu}$, one gets a contribution
\begin{align} \label{delp2}
\Delta p_{\kappa} [S^3]\equiv p_{\kappa}[S^3]-p_{\kappa}[S^2]
\simeq \frac{C_{BS^3}}{12m^2} S^\mu S^\nu S^\lambda
\Bigg[\frac{1}{u}\big(\epsilon_{\alpha\beta\kappa\mu}D_\lambda  
R^{\alpha\beta}_{\quad\delta \nu} u^{\delta}
+\epsilon_{\alpha\beta\gamma\mu} D_\lambda 
R^{\alpha\beta}_{\quad\kappa \nu} u^{\gamma}\big)
\nn\\
-\frac{1}{u^3} \epsilon_{\alpha\beta\gamma\mu} D_\lambda  
R^{\alpha\beta}_{\quad\delta \nu} u^{\gamma} u^{\delta} u_{\kappa}\Bigg].
\end{align}

Yet here with a tensor product of 3 spin vectors, after trading 2 spin vectors by a 
contraction of 2 spin tensors (using again eq.~(4.8) of \cite{Levi:2015msa}), which is 
independent of $u^{\mu}$, the dependence in $u^{\mu}$ of a third spin vector in the product 
still has to be considered. This dependence in fact gives rise to an additional contribution
to the correction in eq.~\eqref{delp2}
\begin{align} \label{delp2add}
\Delta p_{\kappa}^{\text{extra}} [S^3]
\simeq \frac{C_{BS^3}}{6m^2} \frac{D_\lambda B_{\mu\nu}}{u^2} S^\mu S^\nu 
\Big[*S_{\kappa}^{\,\,\lambda}-S^\lambda \frac{u_\kappa}{u}\Big],
\end{align}
where we use here the dual spin tensor defined by, 
\be
*S_{\alpha\beta}\equiv\frac{1}{2}\epsilon_{\alpha\beta\mu\nu}S^{\mu\nu},
\ee
as introduced in \cite{Levi:2015msa}. However, as it turns out the additional correction in 
eq.~\eqref{delp2add} will not contribute to the present sector.

These corrections should be implemented first in the minimal coupling part of the effective 
action of a spinning particle, which is recast in the form \cite{Levi:2015msa}:
\be\label{L_S}
L_{S}=-\frac{1}{2}\hat{S}_{ab}\hat{\Omega}^{ab}_{\text{flat}}
-\frac{1}{2}\hat{S}_{ab} \omega_\mu^{ab} u^{\mu}
-\frac{\hat{S}_{ab}p^b}{p^2}\frac{Dp^a}{D\sigma},
\ee
with lowercase Latin indices for the locally flat frame, and where the Ricci rotation 
coefficients, $\omega_\mu^{ab}$, are used. The new couplings arise from substituting in the 
linear momentum in the canonical gauge into the linear-in-spin couplings, and into the extra 
term that appears last in eq.~\eqref{L_S}, which as noted in section \ref{theory} stands for 
the Thomas precession and was related in \cite{Levi:2015msa} to the gauge of the 
rotational DOFs. 

It is important to stress that the issue here is not about going from a covariant gauge to a 
``non-covariant'' gauge, rather it is about going from the spin-independent approximation of 
the linear momentum to its spin-dependent completion.

From eq.~\eqref{L_S} we obtain the following terms that yield new higher-order in spin 
couplings \cite{Levi:2015msa,Levi:2019kgk}:
\be \label{stos3s4}
L_{S\to S^3, S^4}= 
 \omega_\mu^{ij} u^\mu \frac{\hat{S}_{ik} p^k p_j}{p\left(p+p^0\right)}
-\omega_\mu^{0i} u^\mu \frac{\hat{S}_{ij}p^j}{p}
+\frac{\hat{S}_{ij}p^i \dot{p}^j}{p\left(p+p^0\right)},
\ee
where all the indices are in the locally flat frame, and the canonical gauge is applied.
The first two terms in eq.~\eqref{stos3s4} yield new two-graviton couplings, and the last 
term yields new one-graviton couplings with higher-order time derivatives. Plugging in the 
corrections to the linear momentum from eqs.~\eqref{delp1}, \eqref{delp2} in 
eq.~\eqref{stos3s4} to linear order yields new worldline-graviton couplings that are cubic  
and quartic in the spin, respectively. As noted above, plugging in eq.~\eqref{stos3s4} the 
additional correction in eq.~\eqref{delp2add} is found not to contribute to the present 
sector.

As for the new couplings that are cubic in the spin, these can be found in
\cite{Levi:2019kgk}. The new couplings that are quartic in the spin have the following 
Feynman rules
\begin{align}
\parbox{12mm}{\includegraphics[scale=0.6]{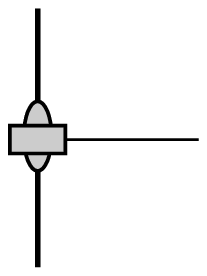}}
 =& \int dt \,\, -\frac{C_{(\text{BS}^3)}}{6m^3}
 \Bigg[S_iS_j\phi_{,ijk}\Big(2\big(S^2a^k
 -\vec{S}\cdot\vec{a} S_k\big)
 +\dot{\vec{S}}\cdot\vec{S} v^k
 -\vec{S}\cdot\vec{v} \dot{S}_k
 \Big)\Bigg],
\end{align}
for the one-graviton coupling, and for a two-graviton coupling
\begin{align}
\parbox{12mm}{\includegraphics[scale=0.6]{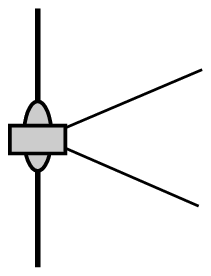}}
 =& \int dt \,\,
 \frac{C_{\text{BS}^3}}{3m^3}  S_iS_jS_kS_l
 \Big[\phi_{,ijk}\phi_{,l}-\delta_{kl}\phi_{,ijm}\phi_{,m} \Big],
\end{align} 
where a gray rectangle mounted on an oval gray blob represents this new type of
``composite'' quartic-in-spin worldline couplings. Further KK field couplings of this type
enter beyond the 5PN order. The Wilson coefficients in these new rules for quartic-in-spin 
couplings indicate that these cannot be identified with the elementary hexadecapole couplings.

\begin{figure}[t]
\centering
\includegraphics[scale=0.7]{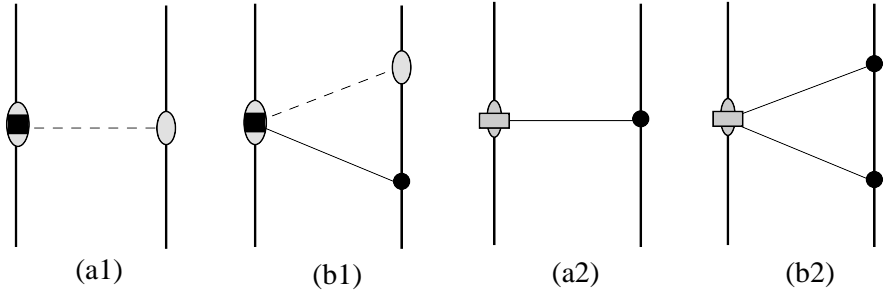}
\caption{The Feynman graphs of one- and two-graviton exchange from composite worldline 
couplings that appear at the NLO quartic-in-spin interaction. 
These graphs with the composite couplings, that are cubic and quartic in the spin, yield 
similar graphs to the corresponding ones with the elementary spin-induced octupole and 
hexadecapole in figure \ref{s4nlo1g}(b1), (c1), and figure \ref{s4nlo2g}(a1), (b1).}
\label{s4nloextra0loop}
\end{figure}

The composite worldline couplings from \cite{Levi:2019kgk} and the new quartic-in-spin 
couplings above give rise to $4$ additional graphs in this sector, as shown in figure
\ref{s4nloextra0loop}, which are similar to those in figure \ref{s4nlo1g} 
(b1), (c1) and in figure \ref{s4nlo2g} (a1), (b1). 
The graphs in figure \ref{s4nloextra0loop} have the following values:
\begin{align}
\text{Fig.~5(a1)}&=\frac{3C_{1(\text{ES}^2)}}{2m_1^2}\frac{G}{r^4}\Bigg[\Big[2S_1^2\big(\vec{S}_2\cdot\vec{a}_1\big)+\big(\dot{\vec{S}}_1\cdot\vec{S}_1\big)\big(\vec{S}_2\cdot\vec{v}_1\big)-\big(\dot{\vec{S}}_1\cdot\vec{S}_2\big)\big(\vec{S}_1\cdot\vec{v}_1\big)\Big]\big(\vec{S}_1\cdot\vec{n}\big)\nonumber\\
&+\Big[2S_1^2\big(\vec{a}_1\cdot\vec{n}\big)+\big(\dot{\vec{S}}_1\cdot\vec{S}_1\big)\big(\vec{v}_1\cdot\vec{n}\big)-\big(\dot{\vec{S}}_1\cdot\vec{n}\big)\big(\vec{S}_1\cdot\vec{v}_1\big)\Big]\Big[\vec{S}_1\cdot\vec{S}_2-5\big(\vec{S}_1\cdot\vec{n}\big)\big(\vec{S}_2\cdot\vec{n}\big)\Big]\nonumber\\
& -2\big(\vec{S}_1\cdot\vec{n}\big)\big(\vec{S}_1\cdot\vec{a}_1\big)\Big(2\big(\vec{S}_1\cdot\vec{S}_2\big)-5\big(\vec{S}_1\cdot\vec{n}\big)\big(\vec{S}_2\cdot\vec{n}\big)\Big)\Bigg],\\
\text{Fig.~5(b1)}&=3C_{1(\text{ES}^2)}\frac{m_2}{m_1^2}\frac{G^2}{r^6}
\Bigg[S_1^2\Big[\vec{S}_1\cdot\vec{S}_2-4\big(\vec{S}_1\cdot\vec{n}\big)\big(\vec{S}_2\cdot\vec{n}\big)\Big]\nonumber\\
&\qquad\qquad\qquad\qquad-\big(\vec{S}_1\cdot\vec{n}\big)^2\Big[2\big(\vec{S}_1\cdot\vec{S}_2\big)-5\big(\vec{S}_1\cdot\vec{n}\big)\big(\vec{S}_2\cdot\vec{n}\big)\Big]\Bigg],\\
\text{Fig.~5(a2)}&= \frac{1}{2}C_{1(\text{BS}^3)}\frac{m_2}{m_1^3}\frac{G}{r^4}\Big[S_1^2-5\big(\vec{S}_1\cdot\vec{n}\big)^2\Big]\nonumber\\
&\quad\times\Big[2S_1^2\big(\vec{a}_1\cdot\vec{n}\big)-2\big(\vec{S}_1\cdot\vec{n}\big)\big(\vec{S}_1\cdot\vec{a}_1\big)+\big(\dot{\vec{S}}_1\cdot\vec{S}_1\big)\big(\vec{v}_1\cdot\vec{n}\big)-\big(\dot{\vec{S}}_1\cdot\vec{n}\big)\big(\vec{S}_1\cdot\vec{v}_1\big)\Big],\\
\text{Fig.~5(b2)}&= C_{1(\text{BS}^3)}\frac{m_2^2}{m_1^3}\frac{G^2}{r^6}
\Big[S_1^4-6S_1^2\big(\vec{S}_1\cdot\vec{n}\big)^2+5\big(\vec{S}_1\cdot\vec{n}\big)^4\Big].
\end{align}

\section{Quadratic-in-curvature worldline couplings}
\label{quadcurv}

In this sector yet a new type of composite coupling, which is quartic in the spin and 
quadratic in the curvature, should be considered: this would arise from plugging the 
correction in 
eq.~\eqref{delp1} back into the leading non-minimal coupling $L_{ES^2}$ in eq.~\eqref{nmc}, 
such that $L_{S^2\to S^4}$, and the resulting coupling would be preceded by the coefficient 
$(C_{\text{ES}^2})^2$. However, this coupling turns out to be power-counted beyond the 5PN 
order, and is thus not relevant here.

\begin{figure}[t]
\centering
\includegraphics[scale=0.7]{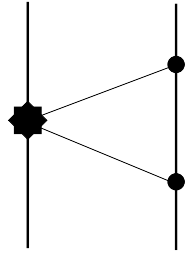}
\caption{The Feynman graph of two-graviton exchange at the NLO quartic-in-spin interaction, 
which originates from the new quadratic-in-curvature operator with the hexadecapole and with 
a new Wilson coefficient.}
\label{s4nlo2gR2}
\end{figure}

We are left then with the elementary coupling of the hexadecapole to the quadratic electric 
operator from eq.~\eqref{frs4E2phi2} as the single contribution to this sector that is 
quadratic in the curvature. The Feynman rule in eq.~\eqref{frs4E2phi2} gives rise to a single 
two-graviton exchange graph, shown in figure \ref{s4nlo2gR2}, the value of which is given by
\begin{align}
\text{Fig.~6}&= 
\frac{C_{1(\text{E}^2\text{S}^4)}m_2^2}{24m_1^3}\frac{G^2}{r^6}
\left[S_1^4-6S_1^2\big(\vec{S}_1\cdot\vec{n}\big)^2+9\big(\vec{S}_1\cdot\vec{n}\big)^4\right]
.&
\end{align}
Notice that this introduces a new Wilson coefficient that appears first in this sector.

\section{Next-to-leading gravitational quartic-in-spin action} 
\label{finalresult}

Putting together the $28=23+4+1=12+11+5$ graph values from sections \ref{Feyncompute}, 
\ref{newfromgauge?}, and \ref{quadcurv}, where the summation also takes into account the 
exchange of particle labels, $1 \leftrightarrow 2$, we get the final effective action for the 
complete quartic-in-spin sector
\begin{equation}
L^{\text{NLO}}_{\text{S}^4}=
L^{\text{NLO}}_{\text{S}_1^2\text{S}_2^2}
+L^{\text{NLO}}_{\text{S}_1^3\text{S}_2}
+L^{\text{NLO}}_{\text{S}_1^4}+(1\leftrightarrow 2),
\end{equation}
where we have
\begin{align}
L^{\text{NLO}}_{\text{S}_1^2\text{S}_2^2}&
=\frac{1}{2}C_{1(\text{ES}^2)}C_{2(\text{ES}^2)}\frac{G}{m_1m_2}\left(\frac{3L_{(1)}}{8r^5}
+\frac{3L_{(2)}}{4r^4}+\frac{L_{(3)}}{2r^3}\right)\nonumber\\*
&\quad-\frac{9}{2}C_{1(\text{ES}^2)}C_{2(\text{ES}^2)}\frac{G^2}{m_1r^6}L_{(4)}
+\frac{1}{2}C_{1(\text{ES}^2)}\frac{G^2}{m_1r^6}L_{(5)},
\end{align}
with the following pieces:
\begin{align}
L_{(1)} =&+10 \big(\vec{S}_1\cdot \vec{v}_1\big) \big[\vec{S}_1\cdot (\vec{v}_1-\vec{v}_2)\big] \big(\vec{S}_2\cdot \vec{n}\big)^2  -10 \big(\vec{S}_2\cdot \vec{v}_2\big)\big[\vec{S}_2\cdot(\vec{v}_1-\vec{v}_2)\big]\big(\vec{S}_1\cdot \vec{n}\big)^2 \nonumber\\ 
& -4 \big[\vec{S}_1\cdot(\vec{v}_1-\vec{v}_2)\big]\big[\vec{S}_2\cdot(\vec{v}_1-\vec{v}_2)\big] \big[\big(\vec{S}_1\cdot \vec{S}_2\big)-5\big(\vec{S}_1\cdot \vec{n}\big)\big(\vec{S}_2\cdot \vec{n}\big)\big] \nonumber\\
& +4 S_1^2 \big(\vec{S}_2\cdot\vec{v}_1\big)^2 -2 S_1^2 \big[\vec{S}_2\cdot(\vec{v}_1+\vec{v}_2)\big] \big(\vec{S}_2\cdot \vec{v}_2\big)  \nonumber\\
& +4 \big(\vec{S}_1\cdot \vec{v}_2\big)^2 {S}_2^2 -2 \big[\vec{S}_1\cdot (\vec{v}_1+\vec{v}_2)\big] \big(\vec{S}_1\cdot \vec{v}_1\big) {S}_2^2 \nonumber\\
& +S_1^2S_2^2\big[5v_1^2+5v_2^2-11 \big(\vec{v}_1\cdot \vec{v}_2\big)-10\big(\vec{v}_1\cdot \vec{n}\big)^2-10\big(\vec{v}_2\cdot \vec{n}\big)^2+15\big(\vec{v}_1\cdot \vec{n}\big) \big(\vec{v}_2\cdot \vec{n}\big)\big]\nonumber\\
& +10 {S}_1^2 \big(\vec{S}_2\cdot \vec{v}_2\big) \big(\vec{S}_2\cdot \vec{n}\big)  \big[(\vec{v}_1+\vec{v}_2)\cdot\vec{n}\big] -10 S_1^2 \big(\vec{S}_2\cdot\vec{v}_1\big)  \big(\vec{S}_2\cdot \vec{n}\big) \big[4\big(\vec{v}_1\cdot\vec{n}\big)-\big(\vec{v}_2\cdot\vec{n}\big)\big]\nonumber\\
& +10 \big(\vec{S}_1\cdot \vec{v}_1\big)  \big(\vec{S}_1\cdot \vec{n}\big) S_2^2 \big[(\vec{v}_1+\vec{v}_2)\cdot \vec{n}\big] -10S_2^2 \big(\vec{S}_1\cdot \vec{v}_2\big)\big(\vec{S}_1\cdot \vec{n}\big)  \big[4\big(\vec{v}_2\cdot \vec{n}\big)-\big(\vec{v}_1\cdot \vec{n}\big)\big] \nonumber\\
& -5\big(\vec{S}_1\cdot \vec{n}\big)^2 S_2^2\big[3v_1^2+5v_2^2-14\big(\vec{v}_2\cdot \vec{n}\big)^2-9\big(\vec{v}_1\cdot \vec{v}_2\big)+7\big(\vec{v}_1\cdot \vec{n}\big) \big(\vec{v}_2\cdot \vec{n}\big)\big]\nonumber\\
& -5S_1^2\big(\vec{S}_2\cdot \vec{n}\big)^2\big[5v_1^2+3v_2^2-14\big(\vec{v}_1\cdot \vec{n}\big)^2-9\big(\vec{v}_1\cdot \vec{v}_2\big)+7\big(\vec{v}_1\cdot \vec{n}\big) \big(\vec{v}_2\cdot \vec{n}\big)\big]\nonumber\\
& +2 \big(\vec{S}_1\cdot \vec{S}_2\big)^2 \big[3v_1^2+3v_2^2-7\big(\vec{v}_1\cdot \vec{v}_2\big)-5\big(\vec{v}_1\cdot \vec{n}\big) \big(\vec{v}_2\cdot \vec{n}\big)\big] \nonumber\\
& +20 \big(\vec{S}_1\cdot \vec{S}_2\big) \big(\vec{S}_1\cdot \vec{v}_1\big) \big(\vec{S}_2\cdot \vec{n}\big) \big[(\vec{v}_1-\vec{v}_2)\cdot \vec{n}\big]-20 \big(\vec{S}_1\cdot \vec{S}_2\big) \big(\vec{S}_1\cdot \vec{v}_2\big) \big(\vec{S}_2\cdot \vec{n}\big) \big(\vec{v}_1\cdot \vec{n}\big) \nonumber\\ 
& -20 \big(\vec{S}_1\cdot \vec{S}_2\big) \big(\vec{S}_1\cdot \vec{n}\big) \big(\vec{S}_2\cdot \vec{v}_2\big) \big[(\vec{v}_1-\vec{v}_2)\cdot \vec{n}\big] -20 \big(\vec{S}_1\cdot\vec{S}_2\big) \big(\vec{S}_1\cdot \vec{n}\big) \big(\vec{S}_2\cdot\vec{v}_1\big) \big(\vec{v}_2\cdot \vec{n}\big) \nonumber\\
& -20\big(\vec{S}_1\cdot \vec{S}_2\big) \big(\vec{S}_1\cdot \vec{n}\big) \big(\vec{S}_2\cdot \vec{n}\big) \big[3v_1^2+3v_2^2-7\big(\vec{v}_1\cdot \vec{v}_2\big)-7\big(\vec{v}_1\cdot \vec{n}\big) \big(\vec{v}_2\cdot \vec{n}\big)\big] \nonumber\\
& -70 \big(\vec{S}_1\cdot \vec{v}_1\big) \big(\vec{S}_1\cdot \vec{n}\big) \big(\vec{S}_2\cdot \vec{n}\big)^2 \big[(\vec{v}_1-\vec{v}_2)\cdot\vec{n}\big] +70 \big(\vec{S}_1\cdot \vec{v}_2\big) \big(\vec{S}_1\cdot \vec{n}\big) \big(\vec{S}_2\cdot \vec{n}\big)^2 \big(\vec{v}_1\cdot \vec{n}\big)\nonumber\\
& +70 \big(\vec{S}_1\cdot \vec{n}\big)^2 
\big(\vec{S}_2\cdot \vec{v}_2\big) \big(\vec{S}_2\cdot \vec{n}\big) \big[(\vec{v}_1-\vec{v}_2)\cdot \vec{n}\big] +70 \big(\vec{S}_1\cdot \vec{n}\big)^2 \big(\vec{S}_2\cdot\vec{v}_1\big) \big(\vec{S}_2\cdot\vec{n}\big)  \big(\vec{v}_2\cdot \vec{n}\big) \nonumber\\
& +35 \big(\vec{S}_1\cdot \vec{n}\big)^2 \big(\vec{S}_2\cdot \vec{n}\big)^2 \big[3v_1^2+3v_2^2-7\big(\vec{v}_1\cdot \vec{v}_2\big)-9\big(\vec{v}_1\cdot \vec{n}\big) \big(\vec{v}_2\cdot \vec{n}\big)\big], 
\end{align}
\begin{align}
L_{(2)}=
& -2 \big(\vec{S}_1\cdot \vec{S}_2\big) \big(\vec{S}_1\cdot \vec{n}\big) \big[2\big(\vec{S}_2\cdot \vec{a}_2\big)+\dot{\vec{S}}_2\cdot\big(2\vec{v}_2-3\vec{v}_1\big)-5 \big(\dot{\vec{S}}_2\cdot \vec{n}\big) \big(\vec{v}_1\cdot \vec{n}\big) \big]\nonumber\\
& +2 \big(\vec{S}_1\cdot \vec{S}_2\big) \big(\vec{S}_2\cdot \vec{n}\big) \big[2\big(\vec{S}_1\cdot \vec{a}_1\big)+\dot{\vec{S}}_1\cdot\big(2\vec{v}_1-3\vec{v}_2\big)-5\big(\dot{\vec{S}}_1\cdot \vec{n}\big) \big(\vec{v}_2\cdot\vec{n}\big)\big]   \nonumber\\
& -2 \big(\vec{S}_1\cdot \vec{S}_2\big) \big[\big(\vec{S}_1\cdot \dot{\vec{S}}_2\big) \big(\vec{v}_1\cdot \vec{n}\big)-\big(\dot{\vec{S}}_1\cdot \vec{S}_2\big) \big(\vec{v}_2\cdot \vec{n}\big)+\big(\vec{S}_1\cdot \vec{v}_1\big) \big(\dot{\vec{S}}_2\cdot \vec{n}\big)-\big(\vec{S}_2\cdot\vec{v}_2\big) \big(\dot{\vec{S}}_1\cdot \vec{n}\big) \big] \nonumber\\
& +2 \big(\dot{\vec{S}}_1\cdot \vec{S}_2\big) \big(\vec{S}_1\cdot \vec{n}\big) \big(\vec{S}_2\cdot \vec{v}_2\big)-2  \big(\vec{S}_1\cdot \dot{\vec{S}}_2\big)  \big(\vec{S}_2\cdot \vec{n}\big)\big(\vec{S}_1\cdot\vec{v}_1\big)\nonumber\\
& -10 \big(\dot{\vec{S}}_1\cdot \vec{S}_2\big) \big(\vec{S}_1\cdot \vec{n}\big)\big(\vec{S}_2\cdot \vec{n}\big) \big(\vec{v}_2\cdot \vec{n}\big) +10 \big(\vec{S}_1\cdot \dot{\vec{S}}_2\big) \big(\vec{S}_1\cdot \vec{n}\big)\big(\vec{S}_2\cdot \vec{n}\big) \big(\vec{v}_1\cdot \vec{n}\big)\nonumber\\
& -2 \big(\vec{S}_1\cdot \dot{\vec{S}}_2\big) \big(\vec{S}_1\cdot \vec{n}\big) \big[\vec{S}_2\cdot \big(2\vec{v}_2-3\vec{v}_1\big)\big]+2 \big(\dot{\vec{S}}_1\cdot \vec{S}_2\big) \big(\vec{S}_2\cdot \vec{n}\big) \big[\vec{S}_1\cdot \big(2\vec{v}_1-3\vec{v}_2\big)\big]  \nonumber\\
&-2 \big[S_1^2-5\big(\vec{S}_1\cdot \vec{n}\big)^2\big]\big[\big(\vec{S}_2\cdot \vec{a}_2\big) \big(\vec{S}_2\cdot\vec{n}\big)+ \big(\dot{\vec{S}}_2\cdot \vec{v}_2\big) \big(\vec{S}_2\cdot \vec{n}\big)+\big(\vec{S}_2\cdot \vec{v}_2\big)\big(\dot{\vec{S}}_2\cdot \vec{n}\big) \big] \nonumber\\
& +2\big[S_2^2-5\big(\vec{S}_2\cdot\vec{n}\big)^2\big]\big[\big(\vec{S}_1\cdot \vec{a}_1\big) \big(\vec{S}_1\cdot \vec{n}\big)+\big(\dot{\vec{S}}_1\cdot\vec{v}_1\big)\big(\vec{S}_1\cdot \vec{n}\big)+\big(\vec{S}_1\cdot \vec{v}_1\big) \big(\dot{\vec{S}}_1\cdot \vec{n}\big)\big] \nonumber\\
& -5\big[S_2^2-3 \big(\vec{S}_2\cdot \vec{n}\big)^2\big]\big[\big(\dot{\vec{S}}_1\cdot\vec{v}_2\big)\big(\vec{S}_1\cdot\vec{n}\big)+\big(\vec{S}_1\cdot \vec{v}_2\big) \big(\dot{\vec{S}}_1\cdot \vec{n}\big)\big] \nonumber\\
& +5 \big[S_1^2-3\big(\vec{S}_1\cdot \vec{n}\big)^2\big] \big[\big(\dot{\vec{S}}_2\cdot \vec{v}_1\big)\big(\vec{S}_2\cdot \vec{n}\big)+\big(\vec{S}_2\cdot\vec{v}_1\big) \big(\dot{\vec{S}}_2\cdot \vec{n}\big)\big]  \nonumber\\
& +5 \big(\vec{S}_1\cdot \vec{n}\big)\big(\dot{\vec{S}}_1\cdot \vec{n}\big) \big[S_2^2 +7\big(\vec{S}_2\cdot \vec{n}\big)^2\big] \big(\vec{v}_2\cdot \vec{n}\big)\nonumber\\
& -5 \big(\vec{S}_2\cdot \vec{n}\big)\big(\dot{\vec{S}}_2\cdot \vec{n}\big) \big[S_1^2 +7\big(\vec{S}_1\cdot \vec{n}\big)^2\big] \big(\vec{v}_1\cdot \vec{n}\big) \nonumber\\
& +8 \big(\vec{S}_1\cdot \vec{v}_2\big) \big(\vec{S}_1\cdot \vec{n}\big) \big(\dot{\vec{S}}_2\cdot \vec{S}_2\big)-8  \big(\vec{S}_2\cdot\vec{v}_1\big) \big(\vec{S}_2\cdot \vec{n}\big) \big(\dot{\vec{S}}_1\cdot \vec{S}_1\big) \nonumber\\
& -2 \big(\vec{S}_1\cdot \vec{v}_1\big) \big(\vec{S}_1\cdot \vec{n}\big) \big[7\big(\dot{\vec{S}}_2\cdot \vec{S}_2\big)-5\big(\vec{S}_2\cdot \vec{n}\big)\big(\dot{\vec{S}}_2\cdot \vec{n}\big)\big] \nonumber\\
& +2 \big(\vec{S}_2\cdot\vec{v}_2\big) \big(\vec{S}_2\cdot \vec{n}\big) \big[7\big(\dot{\vec{S}}_1\cdot \vec{S}_1\big)-5\big(\vec{S}_1\cdot \vec{n}\big) \big(\dot{\vec{S}}_1\cdot \vec{n}\big)\big]\nonumber\\
& +5 \big(\dot{\vec{S}}_1\cdot \vec{S}_1\big) \big[S_2^2-7\big(\vec{S}_2\cdot \vec{n}\big)^2\big]\big(\vec{v}_2\cdot\vec{n}\big)-4 \big(\dot{\vec{S}}_1\cdot \vec{S}_1\big) \big[S_2^2-5\big(\vec{S}_2\cdot \vec{n}\big)^2\big]\big(\vec{v}_1\cdot\vec{n}\big)\nonumber\\
& -5 \big[S_1^2-7\big(\vec{S}_1\cdot \vec{n}\big)^2\big]\big(\dot{\vec{S}}_2\cdot\vec{S}_2\big)\big(\vec{v}_1\cdot\vec{n}\big)+4\big[S_1^2-5\big(\vec{S}_1\cdot \vec{n}\big)^2\big]\big(\dot{\vec{S}}_2\cdot\vec{S}_2\big)\big(\vec{v}_2\cdot\vec{n}\big), 
\end{align}
\begin{align}
L_{(3)}=
& +3 \big(\dot{\vec{S}}_1\cdot \dot{\vec{S}}_2\big) \big(\vec{S}_1\cdot \vec{n}\big)
\big(\vec{S}_2\cdot \vec{n}\big)-3 \big(\dot{\vec{S}}_1\cdot \dot{\vec{S}}_2\big) \big(\vec{S}_1\cdot \vec{S}_2\big)+3 \big(\dot{\vec{S}}_1\cdot \vec{n}\big) \big(\dot{\vec{S}}_2\cdot \vec{n}\big)
\big(\vec{S}_1\cdot \vec{S}_2\big)\nonumber\\
& +13 \big(\dot{\vec{S}}_1\cdot \vec{S}_1\big) \big(\dot{\vec{S}}_2\cdot \vec{S}_2\big) -3\big(\dot{\vec{S}}_1\cdot \vec{S}_2\big) \big(\vec{S}_1\cdot \dot{\vec{S}}_2\big) +15\big(\dot{\vec{S}}_1\cdot \vec{n}\big) \big(\dot{\vec{S}}_2\cdot \vec{n}\big) \big(\vec{S}_1\cdot \vec{n}\big)\big(\vec{S}_2\cdot \vec{n}\big) \nonumber\\
& -21 \big(\dot{\vec{S}}_1\cdot \vec{n}\big) \big(\vec{S}_1\cdot \vec{n}\big) \big(\dot{\vec{S}}_2\cdot \vec{S}_2\big)-21 \big(\dot{\vec{S}}_1\cdot \vec{S}_1\big) \big(\dot{\vec{S}}_2\cdot \vec{n}\big)
\big(\vec{S}_2\cdot \vec{n}\big)
\nonumber\\
& +3 \big(\vec{S}_1\cdot \dot{\vec{S}}_2\big) \big(\dot{\vec{S}}_1\cdot \vec{n}\big) 
\big(\vec{S}_2\cdot \vec{n}\big) +3
\big(\dot{\vec{S}}_1\cdot \vec{S}_2\big) \big(\vec{S}_1\cdot \vec{n}\big) \big(\dot{\vec{S}}_2\cdot \vec{n}\big) \nonumber\\
&+2 \big[\dot{{S}}_1^2+\big(\ddot{\vec{S}}_1\cdot \vec{S}_1\big)\big] \big[S_2^2-3\big(\vec{S}_2\cdot \vec{n}\big)^2\big]+2 \big[S_1^2-3\big(\vec{S}_1\cdot\vec{n}\big)^2\big]  \big[\dot{{S}}_2^2+\big(\ddot{\vec{S}}_2\cdot\vec{S}_2\big)\big], 
\end{align}
\begin{align}
L_{(4)}=
&+S_1^2S_2^2-4S_1^2\big(\vec{S}_2\cdot\vec{n}\big)-4S_2^2\big(\vec{S}_1\cdot\vec{n}\big)^2+\big(\vec{S}_1\cdot\vec{S}_2\big)^2 \nonumber\\ &-12\big(\vec{S}_1\cdot\vec{S}_2\big)\big(\vec{S}_1\cdot\vec{n}\big)\big(\vec{S}_2\cdot\vec{n}\big)+24\big(\vec{S}_1\cdot\vec{n}\big)^2\big(\vec{S}_2\cdot\vec{n}\big)^2,
\end{align}
\begin{align}
L_{(5)}=&+30 \big(\vec{S}_1\cdot \vec{n}\big) \big(\vec{S}_1\cdot \vec{S}_2\big) \big(\vec{S}_2\cdot \vec{n}\big)-57 \big(\vec{S}_1\cdot
\vec{n}\big){}^2 \big(\vec{S}_2\cdot \vec{n}\big)^2 \nonumber\\
& -3 \big(\vec{S}_1\cdot \vec{S}_2\big)^2+9 {S}_1^2 \big(\vec{S}_2\cdot
\vec{n}\big){}^2 +{S}_1^2 {S}_2^2,
\end{align}
and
\begin{align}
L^{\text{NLO}}_{\text{S}_1^3S_2}&=C_{1(\text{BS}^3)}\frac{G}{m_1^2}\left(\frac{{L}_{[1]}}{4r^5}+\frac{{L}_{[2]}}{4r^4}+\frac{{L}_{[3]}}{12r^3}\right)\nonumber\\*
&\quad+C_{1(\text{BS}^3)}\frac{G^2}{m_1r^6}L_{[4]}+C_{1(\text{BS}^3)}\frac{m_2G^2}{m_1^2r^6}L_{[5]}+C_{1(\text{ES}^2)}\frac{G^2}{m_1r^6}L_{[6]}\nonumber\\*
&\quad +\frac{3}{2}C_{1(\text{ES}^2)}\frac{G}{m_1^2 r^4}L_{[7]}+3C_{1(\text{ES}^2)}\frac{m_2 G^2}{m_1^2 r^6}L_{[8]},
\end{align}
with the pieces:
\begin{align}
L_{[1]}=
& -2 \big(\vec{S}_1\cdot \vec{S}_2\big)\big(\vec{S}_1\cdot \vec{v}_1\big)^2 +2\big(\vec{S}_1\cdot \vec{S}_2\big) \big(\vec{S}_1\cdot \vec{v}_1\big) \big(\vec{S}_1\cdot \vec{v}_2\big) \nonumber\\
& -30 \big(\vec{S}_1\cdot \vec{S}_2\big) \big(\vec{S}_1\cdot \vec{v}_1\big) \big(\vec{S}_1\cdot \vec{n}\big) \big[\big(\vec{v}_1-\vec{v}_2\big)\cdot \vec{n}\big] \nonumber\\
& +10 \big(\vec{S}_1\cdot \vec{S}_2\big) \big(\vec{S}_1\cdot \vec{v}_2\big) \big(\vec{S}_1\cdot \vec{n}\big) \big[\big(\vec{v}_1-4\vec{v}_2\big)\cdot
\vec{n}\big]  \nonumber\\
&+\big(\vec{S}_1\cdot\vec{S}_2\big)\Big[13(v_1^2+v_2^2)-29\big(\vec{v}_1\cdot \vec{v}_2\big)-20\big(\vec{v}_1\cdot \vec{n}\big)^2-20\big(\vec{v}_2\cdot \vec{n}\big)^2+25\big(\vec{v}_1\cdot \vec{n}\big)\big(\vec{v}_2\cdot \vec{n}\big)\Big]\nonumber\\
&-5\big(\vec{S}_1\cdot\vec{n}\big)^2\Big[13(v_1^2+v_2^2)-29\big(\vec{v}_1\cdot \vec{v}_2\big)-28\big(\vec{v}_1\cdot \vec{n}\big)^2-28\big(\vec{v}_2\cdot \vec{n}\big)^2+35\big(\vec{v}_1\cdot \vec{n}\big)\big(\vec{v}_2\cdot \vec{n}\big)\Big]\nonumber\\
& -15\big[3S_1^2 -7\big(\vec{S}_1\cdot \vec{n}\big)^2\big]\big(\vec{S}_1\cdot \vec{n}\big) \big(\vec{S}_2\cdot \vec{v}_2\big) \big[\big(\vec{v}_1-\vec{v}_2\big)\cdot \vec{n}\big] \nonumber\\
& +5 \big[3S_1^2-7\big(\vec{S}_1\cdot \vec{n}\big)^2\big] \big(\vec{S}_1\cdot \vec{n}\big) \big(\vec{S}_2\cdot\vec{v}_1\big) \big[\big(4\vec{v}_1-5\vec{v}_2\big)\cdot \vec{n}\big]  \nonumber\\
&-15 S_1^2\big(\vec{S}_1\cdot \vec{n}\big) \big(\vec{S}_2\cdot \vec{n}\big)\Big[3(v_1^2+v_2^2)-7\big(\vec{v}_1\cdot \vec{v}_2\big)-7\big(\vec{v}_1\cdot \vec{n}\big) \big(\vec{v}_2\cdot \vec{n}\big)\Big]\nonumber\\
&+35\big(\vec{S}_1\cdot \vec{n}\big)^3 \big(\vec{S}_2\cdot \vec{n}\big)\Big[3(v_1^2+v_2^2)-7\big(\vec{v}_1\cdot \vec{v}_2\big)-9\big(\vec{v}_1\cdot \vec{n}\big) \big(\vec{v}_2\cdot \vec{n}\big)\Big]\nonumber\\
& +50 \big(\vec{S}_1\cdot \vec{v}_1\big)^2 \big(\vec{S}_1\cdot\vec{n}\big) \big(\vec{S}_2\cdot \vec{n}\big) +40 \big(\vec{S}_1\cdot \vec{v}_2\big)^2 \big(\vec{S}_1\cdot \vec{n}\big) \big(\vec{S}_2\cdot \vec{n}\big) \nonumber\\
& -90
\big(\vec{S}_1\cdot \vec{v}_1\big) \big(\vec{S}_1\cdot \vec{v}_2\big) \big(\vec{S}_1\cdot
\vec{n}\big) \big(\vec{S}_2\cdot \vec{n}\big)  \nonumber\\
& +25 \big[S_1^2 -7\big(\vec{S}_1\cdot\vec{n}\big)^2\big]\big(\vec{S}_1\cdot \vec{v}_1\big)\big(\vec{S}_2\cdot \vec{n}\big) \big[\big(\vec{v}_1-\vec{v}_2\big)\cdot \vec{n}\big] \nonumber\\
& - \big[S_1^2-5\big(\vec{S}_1\cdot\vec{n}\big)^2\big] \big(\vec{S}_1\cdot \vec{v}_1\big)\big[\vec{S}_2\cdot\big(17\vec{v}_1-11\vec{v}_2\big)\big] \nonumber\\
& + \big[S_1^2-5\big(\vec{S}_1\cdot\vec{n}\big)^2\big] \big(\vec{S}_1\cdot \vec{v}_2\big) \big[\vec{S}_2\cdot \big(19\vec{v}_1-13\vec{v}_2\big)\big] \nonumber\\
&  -5 \big[S_1^2-7\big(\vec{S}_1\cdot\vec{n}\big)^2\big] \big(\vec{S}_1\cdot \vec{v}_2\big)
\big(\vec{S}_2\cdot \vec{n}\big) \big[\big(7\vec{v}_1-4\vec{v}_2\big)\cdot \vec{n}\big], 
\end{align}
\begin{align}
L_{[2]}=
& +2  \big(\vec{S}_1\cdot \vec{S}_2\big) \big[S_1^2-10\big(\vec{S}_1\cdot\vec{n}\big)^2\big]  \big(\vec{a}_1\cdot \vec{n}\big)-4\big(\vec{S}_1\cdot \vec{S}_2\big) \big[S_1^2-5\big(\vec{S}_1\cdot\vec{n}\big)^2\big]  \big(\vec{a}_2\cdot \vec{n}\big) \nonumber\\
& -3 \big(\vec{S}_1\cdot \vec{v}_1\big) \big[{S}_1^2-5 \big(\vec{S}_1\cdot \vec{n}\big)^2  \big] \big(\dot{\vec{S}}_2\cdot \vec{n}\big)+4 \big(\vec{S}_1\cdot \vec{v}_2\big) \big[S_1^2-5\big(\vec{S}_1\cdot\vec{n}\big)^2\big]\big(\dot{\vec{S}}_2\cdot \vec{n}\big)  \nonumber\\
&  + \big(\dot{\vec{S}}_1\cdot \vec{v}_2\big) \big[S_1^2+5\big(\vec{S}_1\cdot\vec{n}\big)^2\big] \big(\vec{S}_2\cdot
\vec{n}\big)+4 \big(\vec{S}_1\cdot \vec{a}_2\big) \big[S_1^2-5\big(\vec{S}_1\cdot\vec{n}\big)^2\big] \big(\vec{S}_2\cdot \vec{n}\big)  \nonumber\\
&  +\big[11 S_1^2-45\big(\vec{S}_1\cdot \vec{n}\big)^2\big] \big(\dot{\vec{S}}_1\cdot \vec{n}\big) \big(\vec{S}_2\cdot \vec{v}_2\big)+2  \big(\dot{\vec{S}}_1\cdot\vec{S}_2\big) \big[S_1^2-10\big(\vec{S}_1\cdot \vec{n}\big)^2\big] \big(\vec{v}_1\cdot \vec{n}\big) \nonumber\\
& -2 \big[7S_1^2-10\big(\vec{S}_1\cdot \vec{n}\big)^2\big] \big(\vec{S}_1\cdot \vec{n}\big)\big(\vec{S}_2\cdot\vec{a}_1\big)+5 \big[3S_1^2-7\big(\vec{S}_1\cdot \vec{n}\big)^2\big] \big(\vec{S}_1\cdot \vec{n}\big) \big(\dot{\vec{S}}_2\cdot\vec{n}\big) \big(\vec{v}_1\cdot \vec{n}\big)\nonumber\\
& - \big(\vec{S}_1\cdot \dot{\vec{S}}_2\big) \big[S_1^2-5\big(\vec{S}_1\cdot\vec{n}\big)^2\big] \big(\vec{v}_1\cdot
\vec{n}\big)+\big(\dot{\vec{S}}_1\cdot \vec{S}_2\big) \big[S_1^2+5\big(\vec{S}_1\cdot\vec{n}\big)^2\big]  \big(\vec{v}_2\cdot \vec{n}\big) \nonumber\\
& -2 \big(\dot{\vec{S}}_1\cdot \vec{n}\big)  \big[7S_1^2-30\big(\vec{S}_1\cdot \vec{n}\big)^2\big] \big(\vec{S}_2\cdot\vec{v}_1\big) -4 \big(\vec{S}_1\cdot\dot{\vec{S}}_2\big) \big[S_1^2-5\big(\vec{S}_1\cdot\vec{n}\big)^2\big] \big(\vec{v}_2\cdot \vec{n}\big) \nonumber\\
& -5  \big(\dot{\vec{S}}_1\cdot \vec{n}\big)\big[5S_1^2-21\big(\vec{S}_1\cdot\vec{n}\big)^2\big] \big(\vec{S}_2\cdot \vec{n}\big) \big(\vec{v}_2\cdot \vec{n}\big)-\big[3 S_1^2-5 \big(\vec{S}_1\cdot \vec{n}\big)^2\big] \big(\vec{S}_1\cdot \vec{n}\big) \big(\dot{\vec{S}}_2\cdot\vec{v}_1\big) \nonumber\\
& +8 \big(\vec{S}_1\cdot\vec{S}_2\big)  \big[\big(\vec{S}_1\cdot\vec{a}_1\big) + \big(\dot{\vec{S}}_1\cdot \vec{v}_1\big)-5 \big(\dot{\vec{S}}_1\cdot \vec{n}\big)\big(\vec{v}_1\cdot \vec{n}\big)\big]  \big(\vec{S}_1\cdot \vec{n}\big) \nonumber\\
& -10 \big(\vec{S}_1\cdot \vec{S}_2\big) \big[\big(\dot{\vec{S}}_1\cdot \vec{v}_2\big) -\big(\dot{\vec{S}}_1\cdot \vec{n}\big)\big(\vec{v}_2\cdot \vec{n}\big) \big]
\big(\vec{S}_1\cdot \vec{n}\big)  \nonumber\\
&  +2 \big(\vec{S}_1\cdot \vec{S}_2\big) \big(\dot{\vec{S}}_1\cdot \vec{n}\big) \big[\vec{S}_1\cdot \big(4\vec{v}_1-5\vec{v}_2\big)\big]+2 \big(\vec{S}_1\cdot \vec{S}_2\big) \big(\dot{\vec{S}}_1\cdot \vec{S}_1\big) \big[\big(2\vec{v}_1+\vec{v}_2\big)\cdot \vec{n}\big]\nonumber\\
& -2 \big[\big(\vec{S}_1\cdot \dot{\vec{S}}_2\big) -4\big(\dot{\vec{S}}_1\cdot \vec{S}_2\big)\big]\big(\vec{S}_1\cdot \vec{v}_1\big)
\big(\vec{S}_1\cdot \vec{n}\big) \nonumber\\
& -10 \big[\big(\dot{\vec{S}}_1\cdot \vec{S}_2\big)-\big(\dot{\vec{S}}_1\cdot \vec{n}\big)\big(\vec{S}_2\cdot \vec{n}\big) \big] \big(\vec{S}_1\cdot \vec{v}_2\big)
\big(\vec{S}_1\cdot \vec{n}\big) \nonumber\\
& +2 \big(\dot{\vec{S}}_1\cdot \vec{S}_1\big) \big(\vec{S}_1\cdot \vec{n}\big)\big[11\big(\vec{S}_2\cdot \vec{v}_2\big)  + 10\big(\vec{S}_2\cdot \vec{n}\big) \big(\vec{v}_1\cdot \vec{n}\big) -14 \big(\vec{S}_2\cdot\vec{v}_1\big) -25 \big(\vec{S}_2\cdot \vec{n}\big) \big(\vec{v}_2\cdot \vec{n}\big)  \big]\nonumber\\
&  -2 \big(\dot{\vec{S}}_1\cdot \vec{S}_1\big)
\big[\vec{S}_1\cdot \big(2\vec{v}_1-\vec{v}_2\big)\big] \big(\vec{S}_2\cdot \vec{n}\big)-2 S_1^2 \big[\big(\vec{S}_1\cdot\vec{a}_1\big)+\big(\dot{\vec{S}}_1\cdot\vec{v}_1\big)\big] \big(\vec{S}_2\cdot \vec{n}\big)\nonumber\\
& +10 S_1^2 \big[\big(\vec{S}_1\cdot \vec{n}\big) \big(\vec{a}_1\cdot\vec{n}\big) + \big(\dot{\vec{S}}_1\cdot \vec{n}\big)\big(\vec{v}_1\cdot\vec{n}\big)\big]\big(\vec{S}_2\cdot \vec{n}\big), 
\end{align}
\begin{align}
L_{[3]}=
&+\big(\dot{\vec{S}}_1\cdot
\dot{\vec{S}}_2\big) \big[S_1^2-3\big(\vec{S}_1\cdot \vec{n}\big)^2\big]-9 \big(\dot{\vec{S}}_1\cdot \vec{n}\big)
\big[S_1^2-5\big(\vec{S}_1\cdot \vec{n}\big)^2\big] \big(\dot{\vec{S}}_2\cdot \vec{n}\big) \nonumber\\
& -6 \big(\dot{\vec{S}}_1\cdot \vec{n}\big) \big(\vec{S}_1\cdot \dot{\vec{S}}_2\big) \big(\vec{S}_1\cdot
\vec{n}\big)-18 \big(\dot{\vec{S}}_1\cdot \vec{S}_1\big) \big(\dot{\vec{S}}_2\cdot \vec{n}\big) \big(\vec{S}_1\cdot
\vec{n}\big)+2 \big(\dot{\vec{S}}_1\cdot \vec{S}_1\big) \big(\vec{S}_1\cdot \dot{\vec{S}}_2\big),
\end{align}
\begin{align}
L_{[4]}=-3 \big(\vec{S}_1\cdot\vec{S}_2\big) \big[S_1^2-5\big(\vec{S}_1\cdot \vec{n}\big)^2\big] +2 \big[9S_1^2-20\big(\vec{S}_1\cdot \vec{n}\big)^2\big]\big(\vec{S}_1\cdot \vec{n}\big) \big(\vec{S}_2\cdot
\vec{n}\big),
\end{align}
\begin{align}
L_{[5]}=-9 \big(\vec{S}_1\cdot\vec{S}_2\big) \big[S_1^2-5\big(\vec{S}_1\cdot \vec{n}\big)^2 \big]+51 \big(\vec{S}_1\cdot \vec{n}\big) \big(\vec{S}_2\cdot
\vec{n}\big) S_1^2-115 \big(\vec{S}_1\cdot \vec{n}\big){}^3 \big(\vec{S}_2\cdot \vec{n}\big),
\end{align}
\begin{align}
L_{[6]}=-\big(\vec{S}_1\cdot
\vec{S}_2\big) \big[S_1^2-9\big(\vec{S}_1\cdot \vec{n}\big)^2\big]+6 \big(\vec{S}_1\cdot \vec{n}\big)  \big[S_1^2-4\big(\vec{S}_1\cdot \vec{n}\big)^2\big]\big(\vec{S}_2\cdot\vec{n}\big),
\end{align}
\begin{align}
L_{[7]}&=\Big[2S_1^2\big(\vec{S}_2\cdot\vec{a}_1\big)+\big(\dot{\vec{S}}_1\cdot\vec{S}_1\big)\big(\vec{S}_2\cdot\vec{v}_1\big)-\big(\dot{\vec{S}}_1\cdot\vec{S}_2\big)\big(\vec{S}_1\cdot\vec{v}_1\big)\Big]\big(\vec{S}_1\cdot\vec{n}\big)\nonumber\\
&\quad+\Big[2S_1^2\big(\vec{a}_1\cdot\vec{n}\big)+\big(\dot{\vec{S}}_1\cdot\vec{S}_1\big)\big(\vec{v}_1\cdot\vec{n}\big)-\big(\dot{\vec{S}}_1\cdot\vec{n}\big)\big(\vec{S}_1\cdot\vec{v}_1\big)\Big]\Big[\vec{S}_1\cdot\vec{S}_2-5\big(\vec{S}_1\cdot\vec{n}\big)\big(\vec{S}_2\cdot\vec{n}\big)\Big]\nonumber\\
&\quad -2\big(\vec{S}_1\cdot\vec{n}\big)\big(\vec{S}_1\cdot\vec{a}_1\big)\Big(2\big(\vec{S}_1\cdot\vec{S}_2\big)-5\big(\vec{S}_1\cdot\vec{n}\big)\big(\vec{S}_2\cdot\vec{n}\big)\Big),
\end{align}
\begin{align}
L_{[8]}=S_1^2\Big[\vec{S}_1\cdot\vec{S}_2-4\big(\vec{S}_1\cdot\vec{n}\big)\big(\vec{S}_2\cdot\vec{n}\big)\Big]-\big(\vec{S}_1\cdot\vec{n}\big)^2\Big[2\big(\vec{S}_1\cdot\vec{S}_2\big)-5\big(\vec{S}_1\cdot\vec{n}\big)\big(\vec{S}_2\cdot\vec{n}\big)\Big],
\end{align}
as well as
\begin{align}
L^{\text{NLO}}_{\text{S}_1^4}&=C_{1(\text{ES}^4)}\frac{ Gm_2}{m_1^3}\left(\frac{L_{\{1\}}}{16r^5}+\frac{L_{\{2\}}}{8
r^4}+\frac{L_{\{3\}}}{12r^3}\right)
-\frac{1}{8}C_{1(\text{ES}^4)}\frac{G^2m_2}{r^6m_1^2}L_{\{4\}}
-\frac{3}{8}C_{1(\text{ES}^4)}\frac{G^2m_2^2}{r^6m_1^3}L_{\{5\}}
\nonumber\\
&\quad+C_{1(\text{BS}^3)}\frac{G^2m_2}{r^6m_1^2}L_{\{6\}}-\frac{1}{8}C_{1(\text{ES}^2)}^2\frac{G^2m_2}{r^6m_1^2}L_{\{7\}} \nonumber\\*
&\quad +\frac{1}{2}C_{1(\text{BS}^3)}\frac{G m_2}{r^4m_1^3}L_{\{8\}}+C_{1(\text{BS}^3)}\frac{G^2 m_2^2}{r^6 m_1^3}L_{\{9\}}+\frac{C_{1(\text{E}^2\text{S}^4)}}{24}\frac{G^2m_2^2}{r^6m_1^3}L_{\{10\}},
\end{align}
with the pieces:
\begin{align}
L_{\{1\}}=&+12 \big(\vec{S}_1\cdot \vec{v}_1\big) \big(\vec{S}_1\cdot \vec{v}_2\big) \big[S_1^2-5\big(\vec{S}_1\cdot \vec{n}\big)^2\big]-12 \big(\vec{S}_1\cdot
\vec{v}_1\big)^2 \big[S_1^2-5\big(\vec{S}_1\cdot \vec{n}\big)^2\big]\nonumber\\
& +20
\big(\vec{S}_1\cdot \vec{n}\big) \big[3S_1^2-7\big(\vec{S}_1\cdot \vec{n}\big)^2\big]\big[\big(\vec{S}_1\cdot \vec{v}_1\big) \big(\vec{v}_1-\vec{v}_2\big)\cdot \vec{n}-\big(\vec{S}_1\cdot \vec{v}_2\big)\big(\vec{v}_1\cdot \vec{n}\big)\big]\nonumber\\
&  + \big[3 S_1^4-30S_1^2\big(\vec{S}_1\cdot \vec{n}\big)^2+35\big(\vec{S}_1\cdot \vec{n}\big)^4\big] \big[3v_1^2+3v_2^2-7\big(\vec{v}_1\cdot \vec{v}_2\big)\big]\nonumber\\
& -15\big[S_1^4-14S_1^2\big(\vec{S}_1\cdot\vec{n}\big)^2+21\big(\vec{S}_1\cdot\vec{n}\big)^4\big] \big(\vec{v}_1\cdot \vec{n}\big) \big(\vec{v}_2\cdot \vec{n}\big), 
\end{align}
\begin{align}
L_{\{2\}}=&-6\big(\dot{\vec{S}}_1\cdot \vec{n}\big) \big(\vec{S}_1\cdot \vec{v}_2\big) \big[S_1^2-5\big(\vec{S}_1\cdot \vec{n}\big)^2\big]-10
\big(\dot{\vec{S}}_1\cdot \vec{n}\big) \big(\vec{S}_1\cdot \vec{n}\big) \big[3S_1^2-7\big(\vec{S}_1\cdot \vec{n}\big)^2\big]\big(\vec{v}_2\cdot \vec{n}\big) \nonumber\\
& -12 \big(\dot{\vec{S}}_1\cdot
\vec{S}_1\big) \big(\vec{S}_1\cdot \vec{n}\big) \big(\vec{S}_1\cdot \vec{v}_2\big)+6 \big(\dot{\vec{S}}_1\cdot \vec{S}_1\big)
\big[S_1^2-5\big(\vec{S}_1\cdot \vec{n}\big)^2\big]\big(\vec{v}_2\cdot \vec{n}\big) \nonumber\\
& -2 \big(\dot{\vec{S}}_1\cdot \vec{v}_2\big) \big(\vec{S}_1\cdot \vec{n}\big)
\big[3S_1^2-5 \big(\vec{S}_1\cdot \vec{n}\big)^2\big]- S_1^2\big[S_1^2-5\big(\vec{S}_1\cdot \vec{n}\big)^2\big]\big(\vec{a}_1\cdot \vec{n}\big)\nonumber\\
& -2
\big(\vec{S}_1\cdot\vec{a}_1\big) \big(\vec{S}_1\cdot \vec{n}\big) S_1^2,
\end{align}
\begin{align}
L_{\{3\}}=&3 \big(\ddot{\vec{S}}_1\cdot \vec{n}\big) \big(\vec{S}_1\cdot \vec{n}\big) S_1^2- \big[\dot{{S}}_1^2+\big(\ddot{\vec{S}}_1\cdot \vec{S}_1\big)\big] \big[2S_1^2-3\big(\vec{S}_1\cdot \vec{n}\big)^2\big]\nonumber\\
& +12
\big(\dot{\vec{S}}_1\cdot \vec{n}\big) \big(\dot{\vec{S}}_1\cdot \vec{S}_1\big) \big(\vec{S}_1\cdot \vec{n}\big)+3
\big(\dot{\vec{S}}_1\cdot \vec{n}\big)^2 S_1^2-4 \big(\dot{\vec{S}}_1\cdot \vec{S}_1\big)^2,
\end{align}
\begin{align}
L_{\{4\}}=3S_1^4-30S_1^2\big(\vec{S}_1\cdot\vec{n}\big)^2+35\big(\vec{S}_1\cdot\vec{n}\big)^4,
\end{align}
\begin{align}
L_{\{5\}}=95\big(\vec{S}_1\cdot\vec{n}\big)^4-81S_1^2\big(\vec{S}_1\cdot\vec{n}\big)^2+8S_1^4,
\end{align}
\begin{align}
L_{\{6\}}=\big(\vec{S}_1\cdot\vec{n}\big)^2\big[3S_1^2-5(\vec{S}_1\cdot\vec{n})^2\big],
\end{align}
\begin{align}
L_{\{7\}}=\big[S_1^2-3\big(\vec{S}_1\cdot\vec{n}\big)^2\big]^{2},
\end{align}
\begin{align}
L_{\{8\}}&=\Big[2S_1^2\big(\vec{a}_1\cdot\vec{n}\big)-2\big(\vec{S}_1\cdot\vec{n}\big)\big(\vec{S}_1\cdot\vec{a}_1\big)+\big(\dot{\vec{S}}_1\cdot\vec{S}_1\big)\big(\vec{v}_1\cdot\vec{n}\big)-\big(\dot{\vec{S}}_1\cdot\vec{n}\big)\big(\vec{S}_1\cdot\vec{v}_1\big)\Big]\nonumber\\
&\quad\times\Big[S_1^2-5\big(\vec{S}_1\cdot\vec{n}\big)^2\Big],
\end{align}
\begin{align}
L_{\{9\}}=S_1^4-6S_1^2\big(\vec{S}_1\cdot\vec{n}\big)^2+5\big(\vec{S}_1\cdot\vec{n}\big)^4,
\end{align}
\begin{align}
L_{\{10\}}=S_1^4-6S_1^2\big(\vec{S}_1\cdot\vec{n}\big)^2+9\big(\vec{S}_1\cdot\vec{n}\big)^4.
\end{align}

The result has been ordered according to the Wilson coefficients, mass ratios, and the total 
number/order of higher-order time derivatives. The higher-order time derivatives of both the 
velocity and the spin are to be removed following the procedure that was shown in 
\cite{Levi:2014sba} via variable redefinitions. Then the lengthy result will reduce to an 
ordinary action, and will significantly shrink. But before we handle the higher-order time 
derivatives in this sector, we should take into account additional contributions to this 
sector from lower-order redefinitions of variables made at lower-order sectors, similar to 
what was detailed in section 6 of \cite{Levi:2015msa}. In a forthcoming publication we will 
provide the reduced effective action along with other important quantities and observables 
from this sector.

\section{Conclusions} 
\label{lafin}

In this work we derived for the first time the complete NLO gravitational quartic-in-spin 
interaction of generic compact binaries. The derivation built on the EFT of 
gravitating spinning objects introduced in \cite{Levi:2015msa}, and mainly on its recent 
extensions carried out in 
\cite{Levi:2019kgk,Levi:2020uwu}, in which a new type of worldline couplings should be 
considered, and the effective action of a spinning particle should be extended to quadratic 
order in the curvature.
This sector enters at the 5PN order for maximally-spinning compact objects, and together with 
the NNNLO quadratic-in-spin sector studied in \cite{Levi:2020uwu}, provides all conservative 
finite-size spin effects up to this high PN order. 

Following the cubic-in-spin sector studied in \cite{Levi:2019kgk}, a careful intricate 
analysis was required to recover a new type 
of composite worldline couplings that emerge due the fact that the linear momentum can no 
longer be considered independent of the spin at these nonlinear higher-spin sectors.
These new worldline couplings that contribute here are of cubic and quartic order in the 
spin. It is interesting to consider whether these new couplings can be thought of as the 
classical manifestation of the total angular momenta of a composite particle. 

The analysis in this work shows 
clearly that the spin-dependent correction to the linear momentum will have to be taken into 
account at quadratic order as of the NLO quintic-in-spin level, which corresponds to the 
gravitational Compton scattering with a quantum spin $s=5/2$. This may render such a 
derivation impossibly complex, if not ill-defined. This fundamental connection between the 
classical and the quantum theories with higher spin will be clarified in a forthcoming 
publication.

At the loop computational scale, there was no new nor special difficulty in this sector, and 
in fact this demonstrated once again that even-in-spin sectors are much easier to 
handle compared to odd-in-spin ones, a trend that is clear from inspecting table 
\ref{stateoftheart}. Yet, a new conceptual feature in this sector was a first relevant 
operator that is quadratic in the curvature, and entails a new Wilson coefficient. 

In order to find out what is the effect of the contributions with composite worldline 
couplings from section \ref{newfromgauge?}, and of the new operator quadratic in the 
curvature, we will derive in a forthcoming publication the reduced action, the equations of 
motion, the Hamiltonian, the consequent gauge-invariant observables, and further theoretical 
quantities, which will provide self-consistency checks for the validity of the results.

\acknowledgments

ML has received funding from the European Union's Horizon 2020 research and 
innovation programme under the Marie Sk{\l}odowska-Curie grant agreements 
No.~847523 `INTERACTIONS' and No.~764850 `SAGEX', 
and from the Carlsberg Foundation. 
FT is supported by the Knut and Alice Wallenberg Foundation under grant
KAW 2013.0235, and the Ragnar S\"{o}derberg Foundation (Swedish Foundations’ 
Starting Grant).

\bibliographystyle{JHEP}
\bibliography{gwbibtex}

\end{document}